\documentclass{aa}  

\usepackage{enumitem}
\usepackage{footmisc}
\usepackage{lineno}
\usepackage[colorlinks]{hyperref}
\hypersetup{
  colorlinks   = true, 
  urlcolor     = blue, 
  linkcolor    = blue, 
  citecolor   = cyan 
}

\usepackage{graphicx}
\usepackage{txfonts}
\usepackage{lipsum}
\usepackage{subcaption} 
\usepackage{lscape} 
\usepackage{placeins} 

\def\NH{\textsc{NewHorizon}}
\def\NHtwo{\textsc{NewHorizon2}}
\def\VR{\textsc{VELOCIraptor}}
\def\TF{\textsc{TreeFrog}}
\def\VELUGA{\textsc{VELUGA}}

\def\yzics{\textsc{YZiCS}}
\def\ramses{RAMSES}
\def\yomp{RAMSES-yOMP}

\newcommand{\HI}{\textsc{Hi}}

\begin{document}

   \title{The origin of gas stripping of galaxies in group environments}

    \author{
    Jinsu Rhee\inst{1,2,3}\fnmsep\thanks{Email address: jinsu.rhee@gmail.com} \and
    Christophe Pichon\inst{1,7}\thanks{Corresponding author: pichon@iap.fr} \and
    Yohan Dubois\inst{1} \and
    Sukyoung K. Yi\inst{3} \and
    Jongwan Ko\inst{2} \and
    Yun-Kyeong Sheen\inst{2} \and
    San Han\inst{1} \and
    Seyoung Jeon\inst{3} \and
    J. K. Jang\inst{3} \and
    Wonki Lee\inst{3} \and
    Emanuele Contini\inst{3} \and
    Bumhyun Lee\inst{3} \and
    Jaehyun Lee\inst{2} \and
    Katarina Kraljic\inst{4} \and
    S\'{e}bastien Peirani\inst{5,6}
    }

    \institute{
    Institut d’Astrophysique de Paris, Sorbonne Universit\'{e}, CNRS, UMR 7095, 98 bis bd Arago, 75014 Paris, France \and
    Korea Astronomy and Space Science Institute, 776, Daedeokdae-ro, Yuseong-gu, Daejeon 34055, Republic of Korea \and
    Department of Astronomy and Yonsei University Observatory, Yonsei University, Seoul 03722, Republic of Korea \and
    Observatoire Astronomique de Strasbourg, Universit\'e de Strasbourg, CNRS, UMR 7550, F-67000 Strasbourg, France \and
    ILANCE, CNRS – University of Tokyo International Research Laboratory, Kashiwa, Chiba 277-8582, Japan \and
    Kavli IPMU (WPI), UTIAS, The University of Tokyo, Kashiwa, Chiba 277-8583, Japan \and
    Kyung Hee University, Dept. of Astronomy \& Space Science, Yongin-shi, Gyeonggi-do 17104, Republic of Korea
    }

   \date{Submitted July XX, YYYY}

 \abstract
{We investigate how low-mass group environments ($M_{\rm vir} \sim 10^{12-13}\,M_{\odot}$) influence the gas content of their satellite galaxies with $M_* > 10^{7}\,M_{\odot}$ using the \NHtwo\ simulation.
Many satellite galaxies preserve substantial gas reservoirs, yet show signs of outer gas stripping, reminiscent of jellyfish galaxies in clusters.
In contrast, low-mass satellites ($<10^8 \, M_{\odot}$) are largely gas deficient, and some of them undergo gas removal within their host group by external pressure triggered by either galaxy interactions or ram pressure exerted by the hot intragroup medium.
Complete gas removal in these satellite galaxies occurs when the external hydrodynamic pressure exceeds the gravitational restoring force, typically due to stochastic events such as galaxy-galaxy interactions or nearby galactic outflows.
The emergence of a characteristic stellar mass of $10^8 \, M_{\odot}$, which determines the efficiency of gas removal in groups, likely reflects the differing scaling relations of external pressure with halo mass and gravitational restoring force with stellar mass.
While tidal interactions can be a significant cause of gas loss in satellite galaxies, those severe enough to affect the gas content in the central regions typically lead to the complete disruption of the galaxy.
Consequently, gas loss driven by tidal interactions may be underestimated in the studies focusing solely on surviving galaxies.
Group environments, where environmental effects are weaker and satellite galaxies tend to have lower restoring forces due to their low masses, exhibit complex manifestations of gas loss that are not seen in more massive environments such as clusters.}


   \titlerunning{Origin of gas stripping in groups}
   \authorrunning{Rhee et al.}
   \maketitle   

\nolinenumbers
\section{Introduction}
\label{sec:Int}

Gas in galaxies plays a fundamental role in understanding the formation and evolution of galaxies.
The cold phase of gas acts as the primary fuel for star formation in galaxies.
Rather than fully depleting the gas reservoir through star formation (SF), subsequent supernova explosions suppress further SF by driving mass outflows \citep[][]{SS00, Martin05, Kimm15} or inducing turbulence within the galactic gas \citep[][]{NF96, SO12}.
When bursty SF occurs, the associated outflows generate observable X-ray emission in galaxies \citep[e.g.,][]{McCarthy87,Moran99}.
Additionally, gas accreted from the intergalactic medium or circumgalactic medium (CGM) replenishes the cold gas reservoir in galaxies \citep[][]{Keres05,Dekel09, Rubin12}, thus sustaining the galactic baryon cycle.

The galactic baryon cycle can be significantly disrupted by the surrounding environment.
In massive halos (e.g., $M_{\rm vir}>10^{14}M_{\rm \odot}$), galaxies are often cut off from external gas accretion \citep[][]{McGee14}, and hot halo gas that could otherwise accrete onto galaxies can be stripped away \citep[][]{Larson80,Bosch08,Trussler20,Contini25}.
Galaxy-galaxy interactions \citep[][]{Moore96,Smith10} or tidal interactions \citep[][]{Gao04, Limousin09, Smith16} also perturb internal gas distributions within galaxies.
Among environmental mechanisms, ram pressure \citep[][]{GG72} is considered the most significant and dynamically intense process.
During high-velocity encounters between the intracluster medium (ICM) and interstellar medium (ISM) of a galaxy, the resulting pressure---typically $P_{\rm ram} / k_{\rm B} \sim 10^{5-7} {\rm K}\,{\rm cm}^{-3}$ \citep[][]{Bahe15, Jung18}---can cause substantial removal of the ISM.
This stripping is frequently observed in jellyfish galaxies \citep[][]{Chung09, Sheen17, Yoon17, Jaffe18, Yun19}, which have been successfully reproduced in recent theoretical studies \citep[e.g.,][]{Tonnesen10,LeeJ22}.
On cluster scales, it is widely accepted that galaxies lose most of their gas during pericenter passages, where ram pressure reaches its peak \citep[e.g.,][]{Jung18}.
These mechanisms suppress SF in galaxies, resulting in quiescent red galaxies in dense environments \citep[e.g.,][]{Rhee20}.

Although galaxies in cluster environments have been extensively studied, they constitute only a small fraction of the overall galaxy population, and our understanding of environmentally driven gas removal remains incomplete.
Although each galaxy group contains fewer galaxies than a typical cluster, the much larger number of groups implies that, on cosmic scales, a greater fraction of galaxies reside in groups rather than in clusters \citep[e.g.,][]{Eke04,Robotham11}.
Notably, \cite{Lee22} reported asymmetric distributions even in molecular gas within galaxies in a loose group, suggesting that environmental effects in such environments may be more pronounced than previously recognized.
However, these group environments pose significant challenges for both observations and simulations as their low-mass satellite galaxies are inherently faint and often poorly resolved in numerical models.

The Local Group, a representative example of low-mass group halos \citep[$\sim 3-4\,\times10^{12} M_{\odot}$;][]{Benisty22}, has long served as a valuable laboratory for investigating gas removal processes in group scale halos.
\cite{Einasto74} were among the first to highlight the importance of environmental influence on Local Group satellites; they noted a central concentration of elliptical dwarf galaxies inside the Local Group.
The authors proposed that ram pressure from the intragroup medium (IGM) is a likely cause of the observed morphological dependence on distance from the Local Group center.
Several studies have compiled satellite galaxy catalogs for the Local Group \citep{GP09, McConnachie12, Putman21} and demonstrated that gas-poor satellites are preferentially located near the group center \citep[e.g.,][]{Spekkens14}.
This trend is further supported by \cite{Geha06}, who analyzed the SDSS catalog and found that galaxies at small distances tend to be gas deficient.
The SAGA survey \citep[][]{Geha24}, which achieved a higher completeness of host-satellite samples, also reported similar results.
These findings collectively suggest that gas removal remains a key driver of galaxy evolution even in systems with low virial mass.

However, not all satellite galaxies in group environments undergo severe gas removal.
Based on the compilation by \cite{McConnachie12}, when restricting the sample to Local Group satellite galaxies with $M_{*} > 10^{6}\,M_{\odot}$, located within a radius of $<500\,{\rm kpc}$ to any host, and exhibiting high gas fractions ($f_{\rm gas} > 0.1$), four out of five such satellites---LMC, SMC, IC10, and NGC 6822---have a stellar mass of $M_{*} \gtrsim 10^{8}\,M_{\odot}$.
This suggests that gas-rich satellites are predominantly found among more massive systems, while gas deficiency is more common in satellites with $M_{*} < 10^{8}\,M_{\odot}$.
These trends may reflect longer SF quenching timescales for satellites with $M_{*} > 10^{8}\,M_{\odot}$, and conversely, more rapid quenching in lower-mass galaxies in the Local Group and other observed low-mass groups \citep[][]{Wheeler14, Fillingham15, Weisz15, Wetzel15, Geha24}.
Theoretical studies support this picture, proposing that this critical mass scale emerges from the varying effectiveness of ram pressure stripping, where the process is highly effective for low-mass galaxies, while more massive satellites ($M_{*} > 10^{8}\,M_{\odot}$) are comparatively resistant \citep[e.g.,][]{Fillingham15, Fillingham16, Rhee24}.

Beyond the influence of ram pressure, tidal interactions also play a significant role in shaping the gas content of galaxies.
The tidal force exerted by the host group halo can strip material beyond the tidal radius \citep[e.g.,][]{Read06a, Read06b} and may additionally modulate the effectiveness of ram pressure stripping by altering the galactic potential \citep[e.g.,][]{Mayer06}.
The low-velocity dispersions typical of group systems imply more frequent and prolonged encounters among satellite galaxies, potentially enhancing tidal effects.
For example, \cite{Serra13} reported a giant HI tail in the galaxy group HCG44, likely stripped from NGC 3162 via tidal interactions, \citep[e.g.,][]{For19,Lee-waddell19,Wang22}, highlighting the importance of gravitational interactions within groups in regulating the distribution and retention of cold gas.

The dominant physical mechanism responsible for gas removal in group halos remains unclear, as expected from ongoing observational and theoretical uncertainties.
One of the most well-known cases that exemplifies this ambiguity is the origin of the Magellanic Stream.
Since its discovery \citep[][]{Kuilenburg72,WW72,Mathewson74}, the relative contributions of tidal interaction \citep[e.g.,][]{Mathewson87,Moore94,Hammer15} and ram pressure stripping \citep[e.g.,][]{Gardiner94,Putman98,Besla12} have remained actively debated.
It is now evident that a single dominant mechanism may not be necessary; rather, the interplay between tidal forces and ram pressure likely plays a joint role in shaping the Stream \citep[e.g.,][]{Mayer06, Lin23}.

In conclusion, group halos---although subject to weaker environmental forces than clusters---host satellite galaxies with shallower gravitational potentials, often resulting in more complex and diverse environmental phenomena.
Recent cosmological hydrodynamic simulations have increasingly turned their attention to group-scale systems,\footnote{Table 1 in \cite{Pillepich24} provides a summary of such recent studies.} reflecting a growing interest in their role in galaxy evolution.
While a growing body of work has examined gas removal in these environments, many such studies remain focused on low-mass groups analogous to the Local Group.
Moreover, with few exceptions---most notably the Latte simulation \citep[][]{Wetzel16}---these simulations typically feature spatial resolutions ($>100\,{\rm pc}$) that may be insufficient to resolve internal gas dynamics within galaxies.

The \NHtwo\ simulation \citep[][]{Yi24} is a lower-resolution counterpart to the \NH\ simulation \citep[][]{Dubois21}.
The \NH\ simulation has spatial and mass resolutions high enough to capture the small-scale physics in dwarf galaxies, and includes group systems up to $\lesssim  10^{13}\,M_{\odot}$, which makes it a valuable dataset for studying group environments \citep[][]{Rhee24}.
Since \NHtwo\ adopted the initial condition similar to that of \NH, the halo occupations in the two simulations are nearly identical.
The \NHtwo\ simulation was designed to incorporate the chemical evolution of multiple species to study the origin of galaxy structure, while reducing the resolution by one level to offset computational costs.
In addition, it includes a tracer particle implementation based on \cite{Cadiou19}, enabling the tracking of gas evolution within the Eulerian framework used by \NHtwo .
For this work we utilized the \NHtwo\ simulation to investigate the physical origin of gas removal processes in group halos with $M_{\rm vir} \sim 10^{12-13}\,M_{\odot}$.

\section{Data}
\label{sec:Data}

\subsection{Simulation}
\label{sec:Data-Simulation}

\begin{figure*}
\centering
\includegraphics[width=0.95\textwidth]{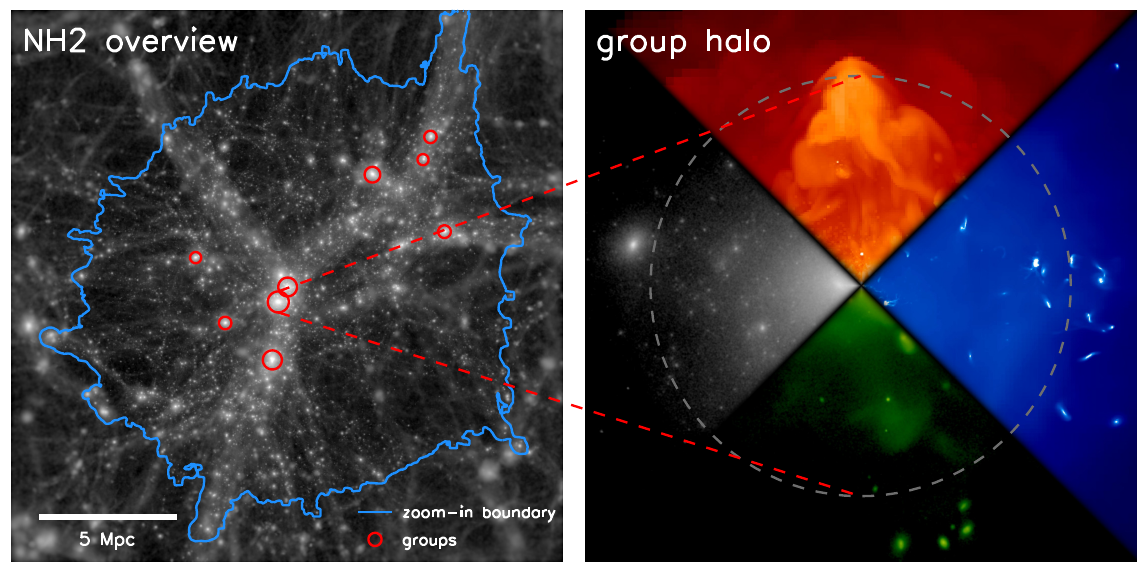}
\caption{
Left panel: Large-scale dark matter density map from the \NHtwo\ simulation surrounding the zoomed-in region.
The boundary of the zoomed-in region is indicated by the blue solid line.
The selected group halos ($M_{\rm vir}>10^{12}\,M_{\odot}$) are marked with the red circles.
Right panel: Enlarged view of one of the group halos.
The different components of the group halo are shown: dark matter in white, gas temperature in red, gas density in blue, and stellar density in green.
The gray dashed circle indicates the region within the virial radius.
}
\label{fig:fig lss}
\end{figure*}

We used the \NHtwo\ simulation \citep[][]{Yi24}, a cosmological zoom-in hydrodynamic simulation, for this analysis.
The simulation was performed using the \yomp\ code, a performance-optimized version of \ramses\ \citep[][]{Teyssier02}, which employs hybrid parallelization via MPI and OpenMP \citep[][]{Han24}.
The same DM only counterpart simulation in a box with a side length of $100\, {\rm cMpc}/{\rm h}$, which was employed in \NH\ \citep[][]{Dubois21}, was used.
The zoomed-in region was selected in the box, similar to the spherical region of $10\, {\rm cMpc}$ in \NH.
However, \NHtwo\ redefines the zoom-in region using a convex hull covering the target particles instead of a spherical region.
This guarantees that the particles in the zoom-in region are less affected by contamination from low-resolution particles entering from outside.
The spatial resolution reaches down to $68 \, {\rm pc}$ in dense regions at $z = 0$.
The mass resolution is $\sim 2 \times 10^{4} M_{\odot}$ for stellar particles and $1.2 \times 10^{6} M_{\odot}$ for dark matter (DM) particles.
Simulation outputs are spaced by $\sim 15\,{\rm Myr}$ time interval, yielding 811 snapshots covering $a=0.0219 - 0.8638$.
\NHtwo\ adopts the following cosmological constants based on the Wilkinson Microwave Anisotropy Probe \citep[][]{Komatsu11}: $\Omega_{m} = 0.272,\  \Omega_{\Lambda} = 0.728,\ \sigma_{8} = 0.81,\ n_{s}=0.967$ and $H_{0} = 70.4 \, {\rm km}\,{\rm s^{-1}}$.

The \NHtwo\ simulation is a lower-resolution counterpart of the \NH\ simulation \citep[][]{Dubois21}.
Both simulations adopt similar astrophysical prescriptions for gas heating and cooling, star formation, and feedback from stars and active galactic nuclei.
However, \NHtwo\ introduces several key differences.
Here, we summarize the astrophysical differences and refer to \cite{Dubois21} and \cite{Yi24} for detailed descriptions.
The main astrophysical additions in \NHtwo\ are the inclusion of Type Ia supernovae (SNe) and stellar winds from AGB stars, aimed at modeling chemical enrichment.
Additionally, the stellar feedback efficiency is enhanced by $50\%$ to mitigate the overcooling of galaxies in the early Universe.
The yields and energy outputs from stellar winds and SNe are computed on-the-fly for each stellar particle, using a simple stellar population model based on Starburst99 \citep[][]{Leitherer99,Leitherer14}.
We adopt the Chabrier initial mass function \citep[][]{Chabrier03} and the Geneva stellar wind model \citep[][]{Schaller92, MM00}.
The SNe Ia rate follows the delay time distribution model from \cite[][]{Maoz12}, assuming a power-law decline over time ($\propto t^{-1}$).
Finally, \NHtwo\ simulation includes tracer particles based on \cite{Cadiou19}, which enables detailed tracking of gas dynamics in the Eulerian framework of the \ramses\ code.

\begin{table}
    \begin{center}
    \caption{\label{tab:table1}Information for the nine group halos}
    \begin{tabular}{cccccc}
    \hline \hline
        Name & $\log M_{\rm vir} / M_{\odot}$\tablefootmark{a} & $R_{\rm vir}$\tablefootmark{b} & $\log M_{*} / M_{\odot}$\tablefootmark{c} & $N_{\rm gal}$ & SFR\tablefootmark{d} \\
        \hline
        G1 & 12.87 & 0.38 & 11.36 & 27 & 0.46 \\
        G2 & 12.74 & 0.35 & 11.14 & 22 & 5.16 \\
        G3 & 12.72 & 0.34 & 11.10 & 19 & 8.78 \\
        G4 & 12.48 & 0.28 & 11.04 & 7 & 3.70 \\
        G5 & 12.21 & 0.23 & 10.75 & 10 & 6.29 \\
        G6 & 12.19 & 0.23 & 10.55 & 3 & 1.89 \\
        G7 & 12.17 & 0.22 & 10.75 & 5 & 2.67 \\
        G8 & 12.04 & 0.20 & 10.53 & 3 & 2.60 \\
        G9 & 12.04 & 0.20 & 10.20 & 4 & 1.45 \\
    \hline
    \end{tabular}
    \end{center}
    \tablefoot{
    \tablefoottext{a}{Virial mass ($M_{\rm 200,c}$) of halos in logarithmic scale}
    \tablefoottext{b}{Virial radius of halos (in Mpc)}
    \tablefoottext{c}{Stellar mass of the brightest group galaxies}
    \tablefoottext{d}{Star formation rate of the brightest group galaxy (in $M_{\odot}\,{\rm yr}^{-1}$)}
    }
\end{table}

\subsection{Group and satellite galaxies}
\label{sec:Data-Galaxies}

We identified halos and galaxies in \NHtwo\ using the six-dimensional friends-of-friends algorithm implemented in \VR\ \citep[][]{Elahi19a}.
For this study, we adopted a performance-enhanced version of \VR\ \citep[][]{Rhee22},\footnote{\href{https://github.com/JinsuRhee/VELOCIraptor-STF}{https://github.com/JinsuRhee/VELOCIraptor-STF}} \footnote{\href{https://github.com/JinsuRhee/NBodylib}{https://github.com/JinsuRhee/NBodylib}} which offers substantial improvements in particle linking via optimized distance metric calculations.
Merger trees for halos and galaxies are constructed using \TF\ \citep[][]{Elahi19b}.
Throughout the post-processing pipeline for halos and galaxies, we make extensive use of the \VELUGA\ software package,\footnote{\href{https://github.com/JinsuRhee/VELUGA}{https://github.com/JinsuRhee/VELUGA}} which is employed to compute global galaxy properties and to repair unstable merger trees.

At the final snapshot ($z=0.158$), the \NHtwo\ simulation contains 3294 halos with $M_{\rm vir} > 10^{9}\,M_{\odot}$.
Here, virial mass is defined as the mass enclosed within a sphere whose average density is 200 times the critical density of the Universe.
From this population, we initially selected ten halos with $M_{\rm vir} > 10^{12}\,M_{\odot}$, targeting the inclusion of Local Group-like systems at the low-mass end of the group halo range.
One halo was excluded due to its proximity to the edge of the zoom-in region and significant contamination ($>5\,\%$).\footnote{Here, contamination refers to the mass fraction of low-resolution DM particles ($M_{\rm DM}>1.2\times10^{6}\,M_{\odot}$) within $R_{\rm vir}$.} 
Galaxies were initially selected with $M_{*} > 10^{7}\,M_{\odot}$, corresponding to approximately 500 stellar particles per galaxy.
While many cosmological simulations adopt a minimum of 100 particles per system to avoid artificial disruption of systems \citep[e.g.,][]{Klypin99}, we adopt a more conservative threshold that ensures robust sampling of galaxy properties while still including low-mass galaxies owing to the high mass resolution of \NHtwo .
In total, 156 galaxies with $M_{*} > 10^{7}\,M_{\odot}$ were identified in $3\,R_{\rm vir}$ of the nine halos.
Of these, 56 galaxies were further excluded due to excessive contamination, broken branches in the merger tree, or being in the first infall stage, which are largely unaffected by group environments.
The final sample thus consists of 100 galaxies, including the nine brightest group galaxies (BGG).
Figure~\ref{fig:fig lss} provides a visual overview of the simulation and the selected halos.
The left panel shows the large-scale structure of the zoom-in volume, with the blue outline marking the boundary of the zoom-in region where full hydrodynamic calculations are performed.
The selected group halos are highlighted with red circles, and one of these is chosen for zoom-in display in the right panel.
The right panel shows four key physical components of the selected group halo: gas temperature (red), gas density (blue), stellar density (green), and dark matter density (white).
The gray circle indicates the virial radius of this halo.
Table~\ref{tab:table1} lists properties of the nine group halos of their virial mass, virial radius, the stellar mass of BGG, the number of satellites and star formation rate (SFR) of BGG.
The host galaxy shows a strong outflow (see the top quadrant on the right panel of the Figure \ref{fig:fig lss}), driven by SN feedback following a starburst triggered by a recent major merger.
This outflow may affect the surrounding satellite galaxies.
We discuss this possibility in Sect.~\ref{sec:Discussion-RP}.

\subsection{Gas properties of satellite galaxies}
\label{sec:Data-Gas}

\begin{figure}
\centering
\includegraphics[width=0.48\textwidth]{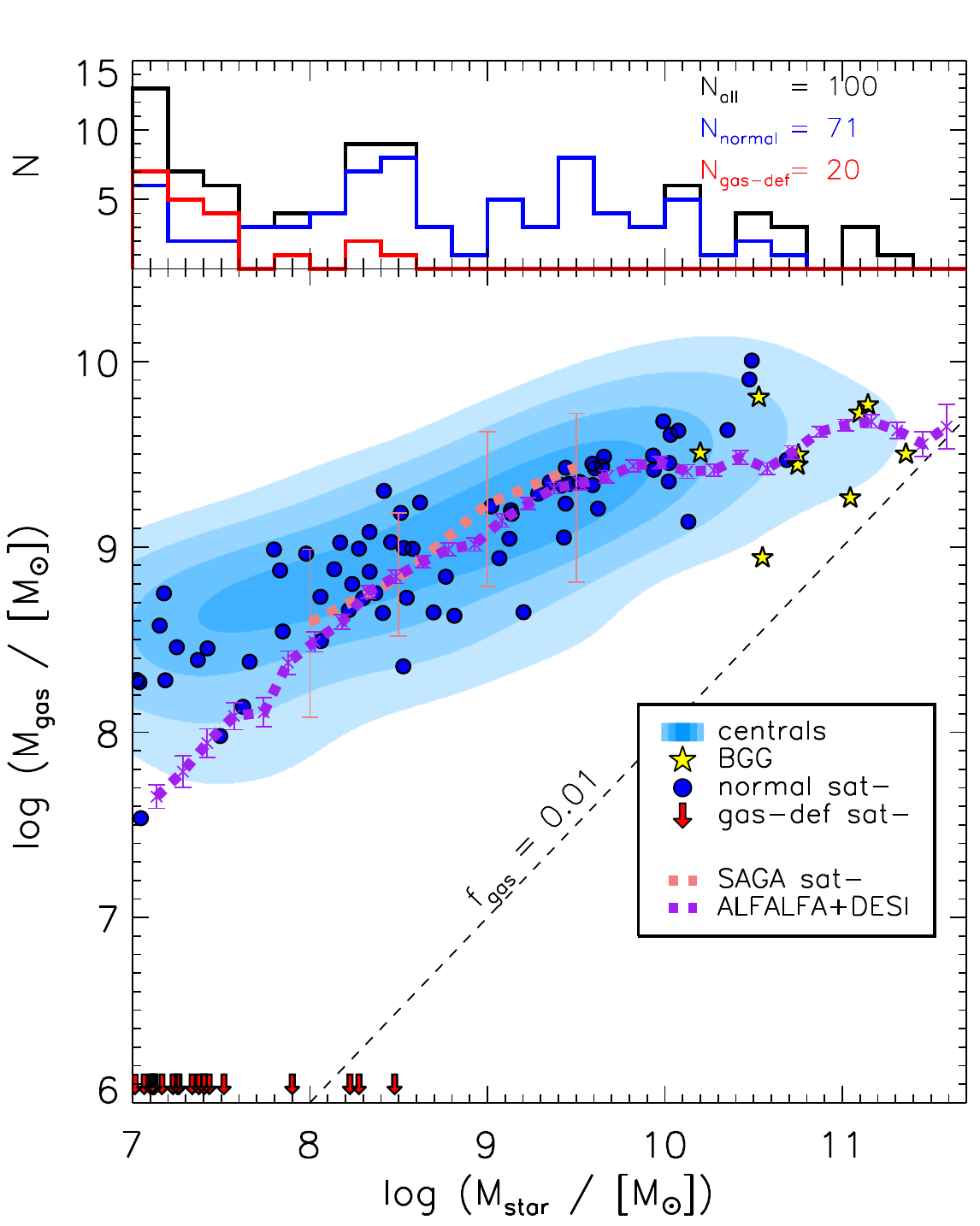}
\caption{
Stellar mass--gas mass scaling relation for satellite galaxies at the final snapshot ($z=0.158$) in the \NHtwo\ simulation.
Satellite galaxies are categorized as either normal ($f_{\rm gas}>0.01$; blue circles) or gas deficient ($f_{\rm gas}<0.01$; downward red arrows).
For comparison, the scaling relation of the \NHtwo\ central galaxies is shown as the blue contours.
Gas-rich normal satellite galaxies closely follow the scaling relation of the central galaxies, while gas-deficient satellites are predominantly low-mass systems.
The stellar mass distribution of the satellite sample is shown as a histogram at the top.
The observed scaling relations from galaxies in the local Universe (purple dotted line) and in low-mass groups (pink dotted line) are overplotted, both showing similar trends to those of the \NHtwo\ central galaxies.
}
\label{fig:fig scaling}
\end{figure}

We classify the gas around galaxies into three categories (ISM, outflowing, and surrounding) based on gravitational boundness, kinematic properties, and metallicity.
First, we compute the gravitational potential for each galaxy using DM, stars, and gas cells within $5\,R_{\rm eff}$, where $R_{\rm eff}$ is the stellar half-mass radius of the galaxy.
For each gas cell, the internal and kinetic energies (using the relative velocity of a gas cell to the host galaxy) are calculated, and those with negative total energy are identified as gravitationally bound.
This bound component is treated as the galaxy's ISM.
The gas mass of a galaxy is defined as the total ISM mass within $R_{\rm eff}$.
This aperture is chosen because it produces gas--stellar mass scaling relations in good agreement with observations (see Fig.~\ref{fig:fig scaling}), although gas components can extend beyond the stellar discs in general.
Gas cells with positive total energy, elevated metallicity, and radially divergent velocities are classified as outflowing.
Specifically, they must satisfy $Z>\bar{Z}-\delta Z$, where $\bar{Z}$ and $\delta Z$ are the mean and standard deviation of the ISM metallicity at the same radius, respectively, under the assumption that outflow gas has the enriched metallicity than surrounding gas, which is likely comparable to ISM.
Divergent motion is identified via $\vec{r} \cdot \vec{v} > 0$, where $\vec{r}$ and $\vec{v}$ are position and velocity vectors relative to the galaxy center.
The central positions and velocities of galaxies are defined as the position and velocity of the center of mass of the stellar particles belonging to each galaxy.
Gas cells that are unbound but do not meet the outflow criteria are classified as ``surrounding gas,'' which likely includes the galactic CGM and the surrounding IGM.
This component is used to estimate the external pressure experienced by galaxies.

We also measure the internal and external gas pressure to assess the hydrodynamic balance between the ISM and surrounding gas during their interaction.
To quantify the retention of internal gas due to gravity, we calculate the volume weighted average of the gravitational potential on the ISM gas cells using all mass components (DM, star, and gas). 
Although the total ISM mass is defined within $R_{\rm eff}$, many satellite galaxies contain ISM gas that extends beyond this radius.
To capture this extended component, we compute internal pressure using ISM gas cells within $3\,R_{\rm eff}$.
We then define $R_{\rm 90, gas}$, the radius enclosing $90\%$ of the ISM gas in $3\,R_{\rm eff}$, as an approximate boundary between the ISM and the surrounding gas.
The gravitational internal pressure, i.e., the restoring pressure exerted by the galaxy's potential on the ISM, is computed for gas cells within $R_{\rm 90, gas}$:
\begin{equation}
    \label{eq:pgv}
    P_{\rm grav} =
    \frac{\frac{1}{2} \sum\limits_{i} \rho_{i} |\Phi_{i}| V_{i}}{\sum\limits_{i} V_{i}}.
\end{equation}
Here $\Phi_{i}$, $\rho_{i}$, and $V_{\rm i}$ are the gravitational potential, density, and volume of the $i$-th gas cell, respectively.
The external pressure is calculated as the volume-weighted mean of the ram pressure from the surrounding gas cells within $2\,R_{\rm 90, gas}$,
\begin{equation}
    \label{eq:prp}
     P_{\rm ext} = 
     \frac{\sum\limits_{i} \rho_{i} |\vec{v}_{i}|^{2} V_{i} }{\sum\limits_{i} V_{i}},
\end{equation}
where $\rho_{i}$, $\vec{v}_{i}$, and $V_{i}$ are the density, galactic-centric velocity, and the volume of $i$-th gas surrounding gas cell, respectively.
These pressure diagnostics are used to quantitatively assess the role of external hydrodynamic forces in driving gas stripping from satellite galaxies.

\section{Results}
\label{sec:Results}

\begin{figure*}
\centering
\includegraphics[width=0.95\textwidth]{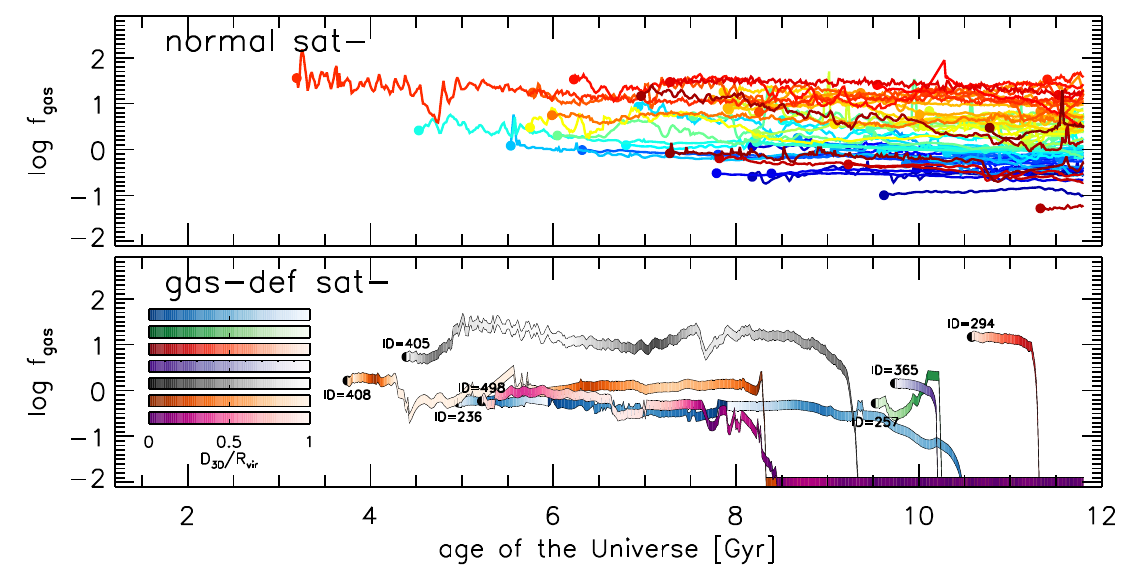}
\caption{
Evolution of gas fractions for normal (top) and gas-deficient satellite galaxies (bottom) over cosmic time.
The colored solid lines in the top panel show the gas fraction of individual normal satellites over cosmic time from their infall to the final snapshot ($z=0.158$).
In the bottom panel, the gas fraction evolution of the seven gas-deficient galaxies that lost their gas inside the host group is shown.
The corresponding galaxy ID for each gas-deficient satellite is also displayed.
In each line, the color intensity reflects the distance to the host group center: bolder colors indicate closer proximity (see color bars on the left side).
Normal galaxies maintain a relatively constant gas fraction, while gas-deficient satellites experience a rapid loss of gas near the host group center.
}
\label{fig:fig gmassevol}
\end{figure*}

\subsection{Gas deficiency of satellites}
\label{sec:Results-Scaling}

We begin by examining the ISM gas--stellar mass scaling relation for the 100 galaxies in our sample, comprising 9 BGGs and 91 satellites (see Fig.~\ref{fig:fig scaling}).
The satellite galaxies are divided into normal ($f_{\rm gas} > 0.01$) and gas-deficient ($f_{\rm gas} < 0.01$) populations, where $f_{\rm gas} = M_{\rm ISM} / M_{*}$.
Among them, 71 are classified as normal (blue circles) and 20 as gas deficient (red downward arrows), while the nine BGGs are shown as yellow stars.
We also compare the relations with central galaxies from all halos in the \NHtwo\ simulation.
Their gas--stellar mass relation is illustrated via blue background contours, showing the $0.5$-, $1$-, $1.5$-, and $2$-$\sigma$ distributions.
We also plot observational scaling relations for dwarf galaxies in the local Universe from \cite{Scholte24}, derived from the Arecibo Legacy Fast ALFA survey \citep[ALFALFA;][]{Giovanelli05, Haynes18} and the Dark Energy Spectroscopic Instrument \citep[DESI;][]{Levi13, Hahn23}, which together extend \HI\ scaling relation of galaxies into the dwarf mass regime (purple dotted line).
For direct comparison with observed satellite populations, we include the gas--stellar mass relation (pink dotted line) derived from the 61 satellites in SAGA DR3 \citep[][]{Geha24,Mao24}.
The upper panel of Fig.~\ref{fig:fig scaling} displays the stellar mass distributions of the full sample, and of the normal and gas-deficient satellite subsamples.

The scaling relations for central galaxies in the \NHtwo\ simulation (blue contours) and in observation (purple dotted line) show overall agreement, though the \NHtwo\ centrals tend to have higher gas mass at the low-mass end.
This offset is at least partly attributed to the ISM definition adopted in this study, which includes all gas components, whereas the observational scaling relations are based on \HI\ gas mass alone.
However, the impact of this difference is expected to be minor, given that molecular gas comprises only a small fraction ($\sim 0.1$) of the \HI\ mass in typical low-mass galaxies \citep[][]{Catinella18}.
In addition, numerical artifacts in the simulation tend to produce artificially extended effective radii for low-mass galaxies, as also seen for \NH\ galaxies in \cite{Martin25}, which leads to an overestimation of the total gas mass.

The normal \NHtwo\ satellite galaxies (red circles) exhibit gas fractions comparable to those of central galaxies, while the gas-deficient population is dominated by low-mass satellites with $M_{*} \lesssim 10^{8}\,M_{\odot}$.
Few satellites are found in the intermediate regime between these two populations, suggesting that a rapid and efficient transition is driven, perhaps due to environmental processes.
These results indicate that, in low-mass group halos ($M_{\rm vir} = 10^{12-13}\,M_{\odot}$), gas stripping is highly efficient, if it occurs, for low-mass satellites and likely operates on a short timescale following infall.

\begin{figure*}
\centering
\includegraphics[width=0.9\textwidth]{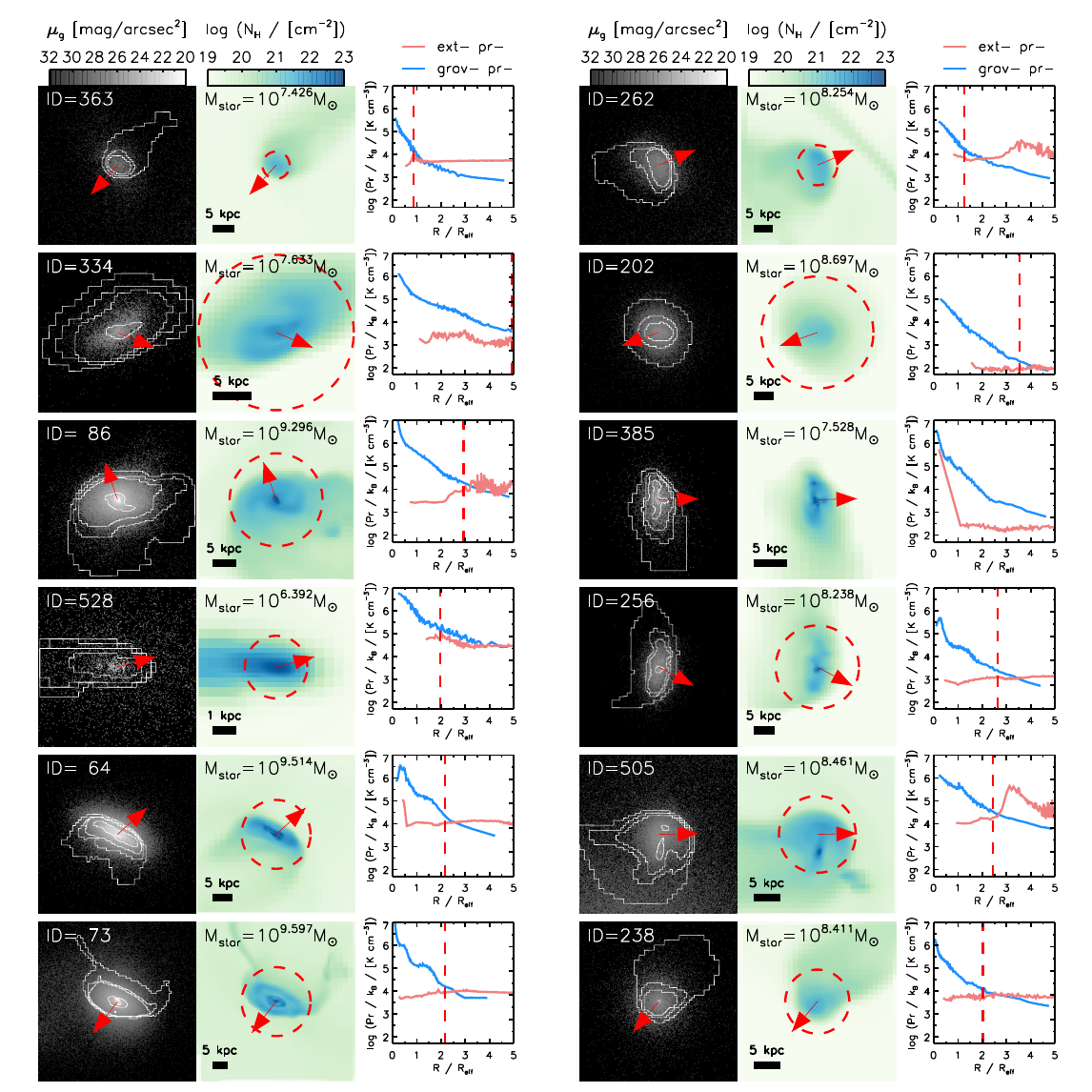}
\caption{
Stellar and gas morphologies of randomly selected normal satellites at their pericenter passages.
Each horizontal triplet corresponds to one galaxy.
Left panel: \textit{g}-band surface brightness of a normal satellite galaxy with gas column density contours.
Middle panel: Gas column density map.
Right panel: Radial profiles of external pressure (pink solid line) and gravitational pressure (blue solid line), measured in spherical shells centered on the galaxy.
In the left and middle panels, the red arrows indicate the direction of the motion of the galaxy.
Galaxy ID and stellar mass are given at the top of the images.
The stripping radius, defined as the location where $P_{\rm ext}=0.5\,P_{\rm grav}$, is marked by the red dashed vertical line in the pressure profile and by the circle in the gas column density map, if it exists.
}
\label{fig:fig himap}
\end{figure*}

The cutoff mass for gas deficiency observed in \NHtwo\ satellites is consistent with recent observational studies.
Satellite galaxies in SAGA DR3 (red solid line) \citep[][]{Geha24} show a similar gas--stellar mass scaling relation to that of \NHtwo\ satellites.
Although the scaling relation from the SAGA sample is likely biased toward gas-rich satellites due to detection limit, it still indicates that when gas is detected, satellites generally retain substantial gas reservoirs, particularly for galaxies with $M_{*} \gtrsim 10^{8}\,M_{\odot}$.
Consistent trends are found in the Local Group: compilations of its satellites \citep[e.g.,][]{GP09, McConnachie12, Putman21} show that only a few satellites---including the Magellanic Clouds--- remain gas rich, and these are all relatively massive ($M_{*} \gtrsim 10^{8}\,M_{\odot}$).
Recent numerical simulations further support this picture, predicting an abundant gas content in satellites above $M_{*} \sim 10^{8}\,M_{\odot}$ \citep[e.g.,][]{Fillingham15, Simpson18, Engler23}.

Figure~\ref{fig:fig gmassevol} shows the evolution of gas fractions for satellite galaxies as a function of cosmic time.
The upper panel presents the individual gas fraction histories of all normal satellite galaxies, with each line representing a single galaxy starting from its infall epoch.
Notably, none of the normal satellites transition into the gas-deficient regime ($f_{\rm gas} < 0.01$) during their post-infall evolution.
This behavior suggests that environmental effects within group halos are insufficiently strong to remove gas from these relatively massive satellites.

Among the 20 gas-deficient satellite galaxies, 12 were already gas poor prior to accretion into their host group through gas loss in filaments \citep[][]{Vulcani21, Castignani22, Jhee22, Yoon25} or in the previous host, as a form of group pre-processing effects \citep[e.g.,][]{Samuel23} or internal feedback process \citep[][]{Jackson21}.
One galaxy became gas deficient as a result of strong internal feedback processes (e.g., supernovae) inside the host group.
Hence, only seven gas-deficient galaxies experienced gas removal within the group environment.
The bottom panel of Fig.~\ref{fig:fig gmassevol} shows the gas fraction evolution of these seven satellites with their galaxy identifier (ID).
Each line is colored to become progressively bolder as each galaxy approaches the group center (see the inset color tables), to highlight their orbital evolution.
The sharp decline in gas fraction for these satellites indicates that their ISM reservoirs are stripped on short timescales, typically coinciding with their closest approach to the group center, as inferred from the color gradients.
This spatial and temporal correspondence strongly supports ram pressure stripping as the dominant mechanism responsible for gas loss in these systems \citep[e.g.,][]{Simpson18, Engler23, Samuel23}.

\subsection{Jellyfish signature of normal satellites}
\label{sec:Results-Normal}

Figures~\ref{fig:fig scaling} and \ref{fig:fig gmassevol} suggest that normal satellite galaxies remain largely unaffected by gas-removal processes in group halos.
Considering that interaction between the hot IGM and the ISM is prevalent within the virial radius of group environments \citep[e.g.,][]{Rhee24}, this raises the question of whether there is observational evidence for such interactions.
To investigate the signatures of IGM-ISM interaction, we selected a random subset of normal satellite galaxies and examined their stellar and gas distributions (Fig.~\ref{fig:fig himap}).
Each galaxy is shown at its first pericenter passage, the epoch when environmental effects are expected to be strongest.
Each set of panels in Fig.~\ref{fig:fig himap} includes the \textit{g}-band stellar surface brightness map (left) and the ISM column density map (middle).
On each stellar surface brightness map, ISM column density contours at $1$, $5$, $10$, and $50 \times 10^{20} \,{\rm cm}^{-2}$ are overlaid, while the red arrows indicate the direction of motion of the galaxies with respect to the host group center.
The right panel in each set displays the radial profiles of gravitational pressure (blue solid line) and external pressure (pink solid line), calculated in each radial shell following the method described in Sect.~\ref{sec:Data-Gas}.
The cut-off radius, defined as where $P_{\rm ext} \sim 0.5\,P_{\rm grav}$, is indicated by a red dashed line in the pressure profile and marked by a circle on the ISM map, providing a visual estimate of the region most vulnerable to external stripping forces.
Galaxy ID and stellar masses are labeled in the upper left corners of the respective maps.

Although these satellites remain gas rich, most exhibit noticeable gas tail morphologies, with the exception of the galaxy with ID=202.\footnote{The absence of the gas tail is intrinsic, not due to a projection effect.}
The tails generally extend opposite to the direction of motion, consistent with the action of external forces.
The cut-off radius, where $P_{\rm ext} \sim 0.5\,P_{\rm grav}$, typically corresponds to the region where gas tails originate.
At the same time, their stellar distributions are unaffected, both suggesting that external pressure plays a central role in shaping these features.
At the cut-off radius, the external pressure remains below $P_{\rm ext} / k_{\rm B} < 10^{5}\,{\rm K}\,{\rm cm}^{-3}$, significantly weaker than typical values observed in cluster centers \citep[$P_{\rm ext} / k_{\rm B} \sim 10^{6}\,{\rm K}\,{\rm cm}^{-3}$;][]{Jung18, Yun19, Boselli21}.
This lower-pressure regime favors the formation of gas tails rather than complete stripping of the ISM \citep[e.g.,][]{Lee20}.

Therefore, although these normal satellite galaxies appear largely unaffected in terms of total gas mass (within $R_{\rm eff}$), they commonly have distorted ISM morphology induced by moderately strong external pressure.
Even under peak conditions, i.e., pericenter passages, the external pressure does not reach the threshold required to strip the ISM, but it is sufficient to disrupt and reshape the gas distribution.
This may also account for the temporary SF quenching observed in satellite galaxies with $M_{*}\gtrsim10^{9}\,M_{\odot}$ during their pericentric passages \citep[][]{Rhee24}.
The duration over which such pressures act is short, owing to the high speed of galaxies near the group centers.
Furthermore, the gas in tails can be accreted to the galaxy rather than stripped when external pressure becomes lower to some extent \citep[e.g.,][]{Zhu24}.
Consequently, the combination of limited pressure strength and short interaction timescales allows normal satellites to retain their gas reservoirs while producing observable morphological disturbances without substantial gas loss.

\subsection{Gas stripping origin}
\label{sec:Results-Def}

\begin{figure*}
\centering
\includegraphics[width=0.9\textwidth]{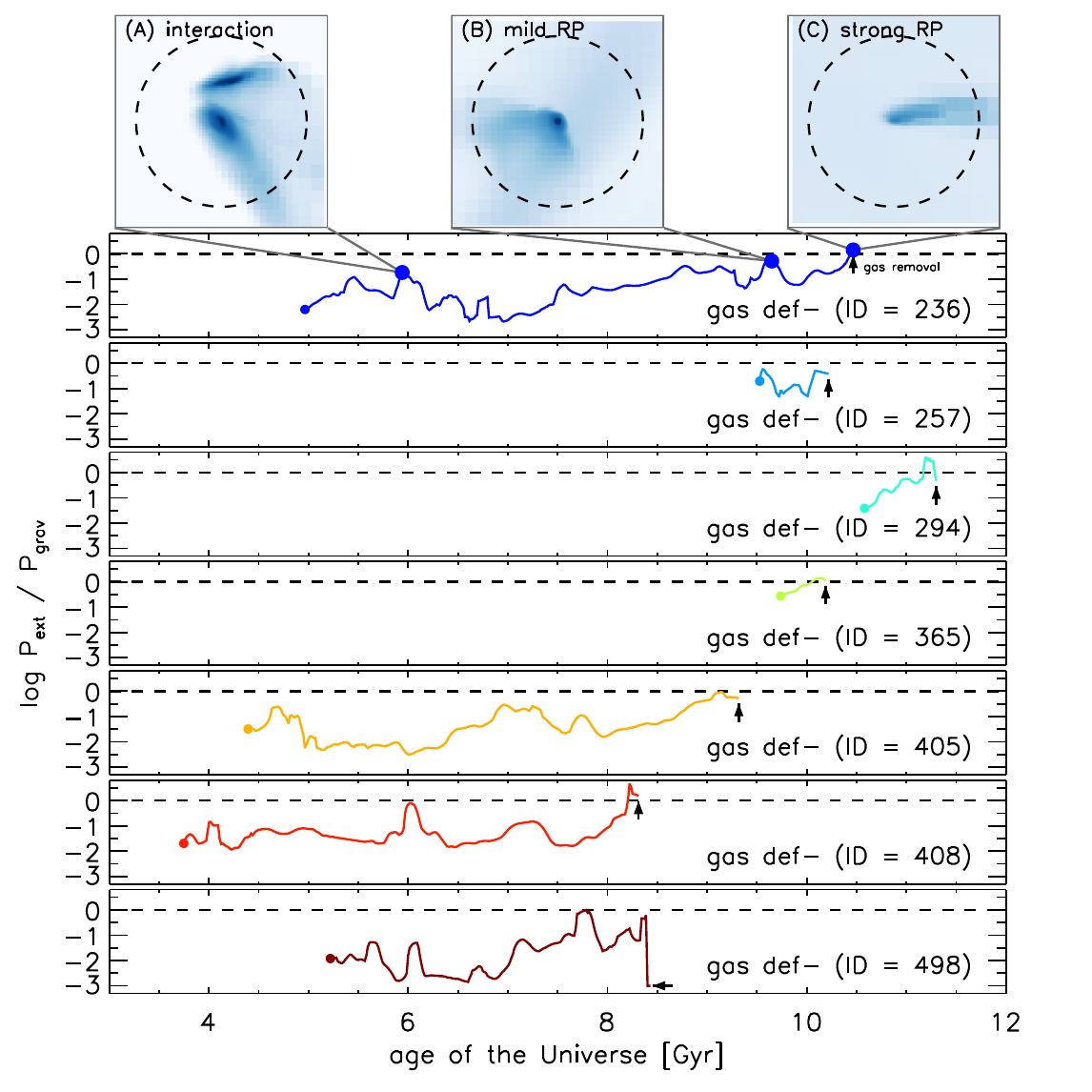}
\caption{
Time evolution of the ratio of external pressure to gravitational pressure for the seven gas-deficient satellites.
Both pressures are computed using Equations~\ref{eq:pgv} and \ref{eq:prp}, and smoothed over a $50\,{\rm Myr}$ window to reduce fluctuations.
The black dashed horizontal lines mark the point at which $P_{\rm ext} = P_{\rm grav}$.
The moment of gas removal for each galaxy is marked with the black arrow.
For the galaxy with ID=236, three distinct sources of strong external pressures are illustrated in the top panels through gas column density maps: (A) fly-by interaction with another galaxy, (B) mild ram pressure, and (C) strong ram pressure.
In each map, the black dashed circle indicates $2\,R_{\rm 90, gas}$, where $R_{\rm 90, gas}$ is the radius enclosing $90\%$ of the ISM gas mass within $3\,R_{\rm eff}$.
}
\label{fig:fig prcompare}
\end{figure*}

As shown in Fig.~\ref{fig:fig gmassevol}, seven gas-deficient satellite galaxies undergo a rapid gas removal process, indicating the presence of a strong external influence.
To investigate the physical origin of this process, we begin by examining the external pressure acting on these galaxies, under the assumption that ram pressure is the most likely driver.
Our analysis in Fig.~\ref{fig:fig himap} demonstrates that the balance between gravitational restoring force and external pressure has strong predictive power in identifying whether gas is retained or stripped.
Building on this, Fig.~\ref{fig:fig prcompare} presents a direct comparison of gravitational and external pressures for the seven gas-deficient satellites, providing further insight into the physical conditions that govern gas stripping in group environments.

\begin{figure*}
\centering
\includegraphics[width=0.95\textwidth]{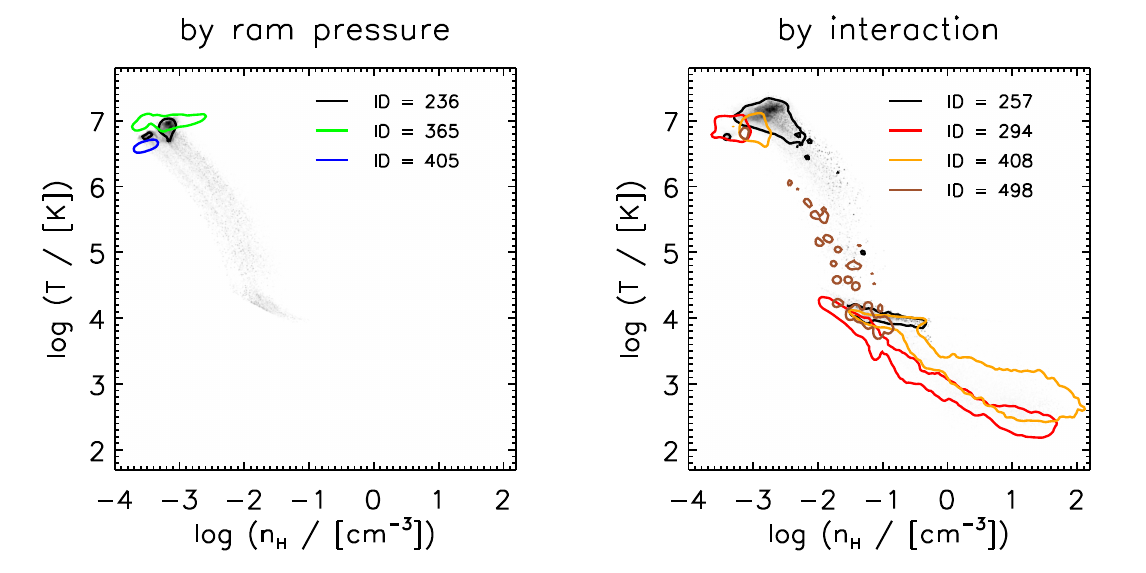}
\caption{
Gas phase distribution of surrounding gas cells of the seven gas-deficient satellites during the last $200\,{\rm Myr}$ before the gas removal.
The distributions are weighted by the external pressure contributed by individual surrounding gas cells.
The satellites are grouped into two categories based on the properties of the surrounding gas: those embedded in hot and low-density gas (left panel), and those surrounded by cold and dense gas (right panel).
In each panel, the $1\,\sigma$ distributions of the surrounding gas in the phase diagram for individual galaxies are shown in different colors.
The background density map displays the full surrounding gas phase distribution for a single satellite in each case (shown in black) as a reference.
Gas-deficient satellites in the left panel are likely stripped by classical ram pressure, whereas those in the right panel are affected by interactions with nearby cold and dense gas structures.
}
\label{fig:fig igmphase}
\end{figure*}

\begin{figure}

\includegraphics[width=0.45\textwidth]{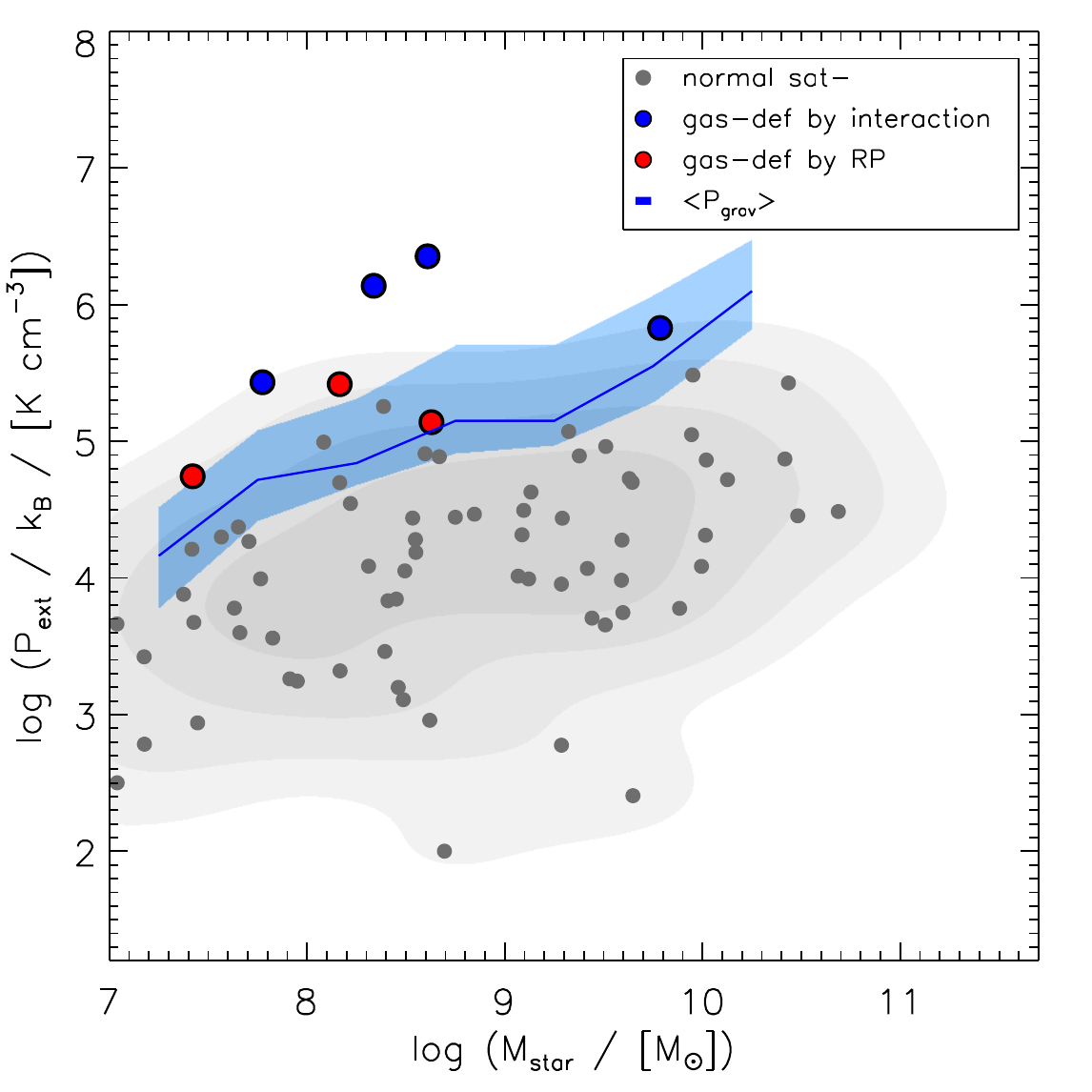}
\caption{
Maximum external pressure vs. stellar mass for satellite galaxies in the \NHtwo\ simulation.
Each gray circle represents the maximum external pressure experienced by an individual satellite, while the background contours indicate the overall distribution.
The seven gas-deficient galaxies, those stripped by galaxy interactions (blue circles) and by ram pressure (red circles), are highlighted.
The blue solid line and shaded region respectively represent the median and quartile ranges of gravitational pressure for satellite galaxies within each stellar mass bin when they are exposed to the maximum external pressure.
Gas-deficient satellites experienced external pressure comparable to or exceeding their gravitational pressure, whereas the majority of satellites remain unaffected due to weaker external pressures.
}
\label{fig:fig pressure}
\end{figure}

In Fig.~\ref{fig:fig prcompare}, the ratio of external pressure ($P_{\rm ext}$) to gravitational pressure ($P_{\rm grav}$) for each individual galaxy is shown as a function of the age of the Universe.
Here, $P_{\rm ext}$ and $P_{\rm grav}$ for each galaxy are measured using Equations~\ref{eq:pgv} and \ref{eq:prp}, and their time evolution is smoothed using a $50\,{\rm Myr}$ time window to reduce their stochastic noise.
The gas removal epoch of the galaxies is shown with the black arrow. 
Similarly, each panel in the figure shows the evolution of a single galaxy from its infall epoch to the moment of gas removal.
The black dashed line in each panel marks the point where $P_{\rm ext} = P_{\rm grav}$.

In all panels of Fig.~\ref{fig:fig prcompare} from top to bottom, the seven gas-deficient galaxies become stripped of their ISM when $P_{\rm ext} \gtrsim P_{\rm grav}$, indicating that external hydrodynamic pressure is the primary driver of their gas loss.
However, not all episodes of elevated external pressure lead to complete gas stripping, and we identify physical pathways through which external pressure can arise.
To illustrate this, we present three representative cases of external pressure evolution for Galaxy with ID=236, with corresponding gas distributions shown in the top three panels.
In Panel-(A) of Fig.~\ref{fig:fig prcompare}, a fly-by interaction with another galaxy generates strong external pressure.
Here, the extended gas envelope of the passing galaxy induces the high external pressure, temporarily elevating $P_{\rm ext}$ without leading to full stripping.
In Panel-(B), a case of mild ram pressure yields external pressure comparable to the gravitational restoring force.
This results in stripping outer gas layers, while the central gas is compressed and retained.
The resulting concentration of ISM enhances the restoring force of the galaxy, allowing it to preserve most of its gas despite displaying a jellyfish-like morphology.
In contrast, Panel-(C) shows a case of strong ram pressure, where the external force significantly exceeds the restoring force.
Under these conditions, the galaxy undergoes complete gas stripping.

Therefore, satellite galaxies are subject to various types of external pressure.
In some cases, the external pressure is insufficient to fully strip the ISM, resulting in jellyfish morphologies with retained central gas (seen as Fig.~\ref{fig:fig himap}).
In contrast, stronger external pressure can overcome the gravitational restoring force, leading to complete gas removal.
These findings naturally raise the question of under what conditions the external pressure becomes comparable to or exceeds the internal gravitational restoring pressure.
Understanding the physical origins and timing of such conditions is essential for predicting when and how galaxies undergo environmental quenching.

To investigate the origin of strong external pressure observed in the seven gas-deficient satellite galaxies in Fig.~\ref{fig:fig prcompare}, we examine each gas-deficient galaxy at the epoch of its gas removal.
Figure~\ref{fig:fig igmphase} presents the phase diagrams of the surrounding gas for each of the seven galaxies during the final $200\,{\rm Myr}$ before gas removal.
In each phase diagram, gas cells are weighted by the magnitude of their external pressure, thereby emphasizing the thermodynamic conditions associated with strong external pressure.
The 1-$\sigma$ distributions of surrounding gas cells for each individual galaxy are shown in different colors, while the black contour and background distribution represent the distribution for a representative comparison sample.

We identify two distinct sources of strong external pressure responsible for gas removal in the seven satellites.
In the left panel of Fig.~\ref{fig:fig igmphase}, galaxies experience strong external pressure from surrounding gas cells that are hot and low in density.
These cases are consistent with the canonical form of ram pressure, driven by the interaction between hot gas and galactic ISM.
In contrast, the four galaxies in the right panel are embedded in cold, dense gas environments, despite the presence of some hot gas.
In these systems, the dominant external pressure arises from cold high-density gas, which is likely associated with the extended disk of another galaxy or a dense gas clump.
Individual analyses (not shown here) confirm that these galaxies encounter cold gas streams or extended gaseous structures from neighboring galaxies, which exert localized but intense ram pressure, leading to rapid and efficient stripping of their ISM. 
Therefore, unlike the conventional understanding of ram pressure in cluster galaxies, group galaxies may experience external pressure from different origins: not only from hot gas (as in classical ram pressure), but also from cold gas streams or extended disks \citep[e.g.,][]{Marasco16}.
However, two galaxies (ID=257 and ID=408) also undergo severe mass loss during the interactions, implying that gravitational tides from neighbor galaxies as well as hydrodynamic pressure are co-acting for these galaxies.

Although many satellite galaxies are exposed to external pressure, only a subset undergo significant gas stripping.
Figure~\ref{fig:fig pressure} presents the maximum external pressure experienced by each satellite galaxy over its evolutionary history.
Gray circles denote gas-rich normal satellites, with their distribution indicated by background contours.
In contrast, blue and red circles represent the gas-deficient satellites.
To assess whether the maximum external pressure exceeded the internal gravitational restoring force, we overlay the median gravitational pressure of the satellite population when the maximum external pressure is exerted as the blue solid line, with the interquartile range shown as a shaded blue region.
This comparison allows us to identify which galaxies experienced external pressure sufficient to overcome their self-gravity, resulting in gas loss.
We note that, in the plot, the stellar masses correspond not to the present values, but to those at the time when the external pressure was at its peak.
As a result, the galaxies generally appear more massive than they are at the final snapshot, especially those without gas, since they have experienced continuous mass loss over time without SF.

\begin{figure*}
\centering
\includegraphics[width=0.88\textwidth]{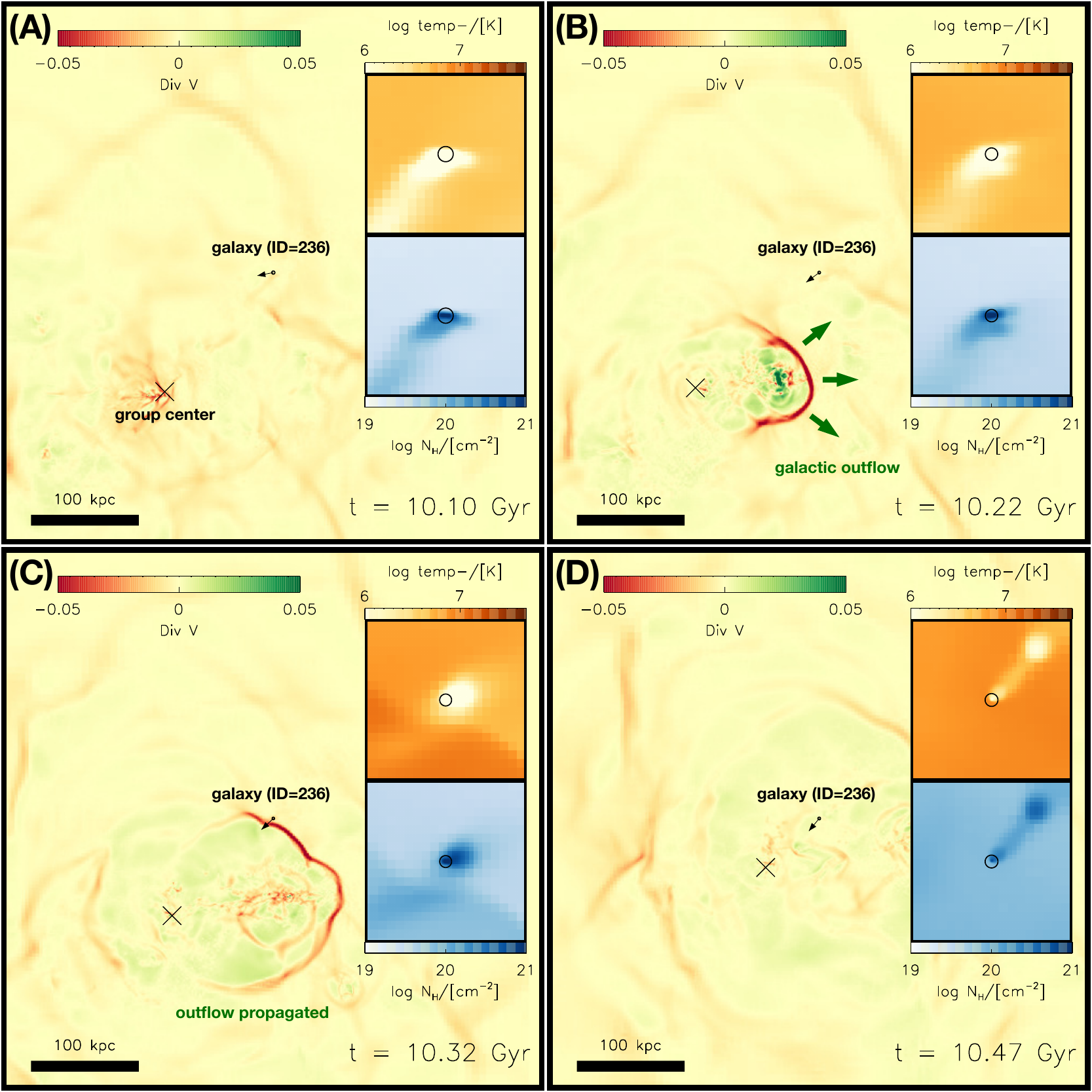}
\caption{
Time sequence of gas stripping in Galaxy ID=236, shown from Panels (A) to (D).
In each panel, the background color map displays the gas velocity divergence, highlighting shock fronts as thin layers of converging flows.
The target galaxy is located at the center, and its direction of motion is shown with the black arrow.
The host group center is marked by a cross ($\times$).
Two images in each panel provide zoomed-in views of the target galaxy, showing gas temperature (top) and column density (bottom).
In Panel (B), a shock generated by a galactic outflow from a nearby galaxy begins to propagate toward the target galaxy.
In Panel (C), the two insets show the discontinuity in  temperature and in density of the surrounding gas along the shock front.
The shock interacts with the target galaxy, resulting in strong external pressure on the galaxy.
}
\label{fig:fig rporigin}
\end{figure*}

Gas-rich normal satellite galaxies retain their ISM because the external pressure that they experience remains below their gravitational restoring pressure, allowing them to preserve gas over an extended timescale.
In contrast, the seven gas-deficient satellite galaxies exhibit exceptionally high peak external pressure, significantly exceeding the median gravitational pressure of galaxies in the same stellar mass range.
These results indicate that gas stripping requires external pressure to surpass the internal restoring force by a considerable margin.
Moreover, such extreme external pressure appears to arise from stochastic physical processes, including close galaxy-galaxy interactions, encounters with dense gas clumps, or exposure to outflows from neighboring systems.

\section{Discussion}

\subsection{Origin of strong and anomalous external pressure}
\label{sec:Discussion-RP}

Figures.~\ref{fig:fig igmphase} and~\ref{fig:fig pressure} show that three satellite galaxies lose their ISM gas due to strong ram pressure.
However, the origin of this enhanced ram pressure is not immediately clear, as other satellites with comparable stellar masses experience significantly weaker external pressure.
A detailed analysis of individual gas-deficient galaxies reveals a specific physical process responsible for generating strong external pressure.

Figure~\ref{fig:fig rporigin} presents the evolution of galaxy with ID=236, which becomes gas deficient as a result of strong ram pressure stripping.
Panels (A) through (D) track the motion of the galaxy, with the galaxy centered in each panel, and the host group's center marked by an X symbol.
The direction of motion of the target galaxy is denoted by the black arrow.
In each panel, the background color map shows the gas velocity divergence, calculated from local velocity gradients across neighboring cells.
Shock fronts, which exert additional ram pressure on the satellites, are highlighted as thin layers of converging flows \citep[$\nabla \cdot \vec{v} < 0$;][]{Ryu03}.
Each panel includes two insets showing zoom-in views of the galaxy within $30\,{\rm kpc}$ region: the upper inset displays the gas temperature map, while the lower inset shows the gas column density.

\begin{figure*}
\centering
\includegraphics[width=0.95\textwidth]{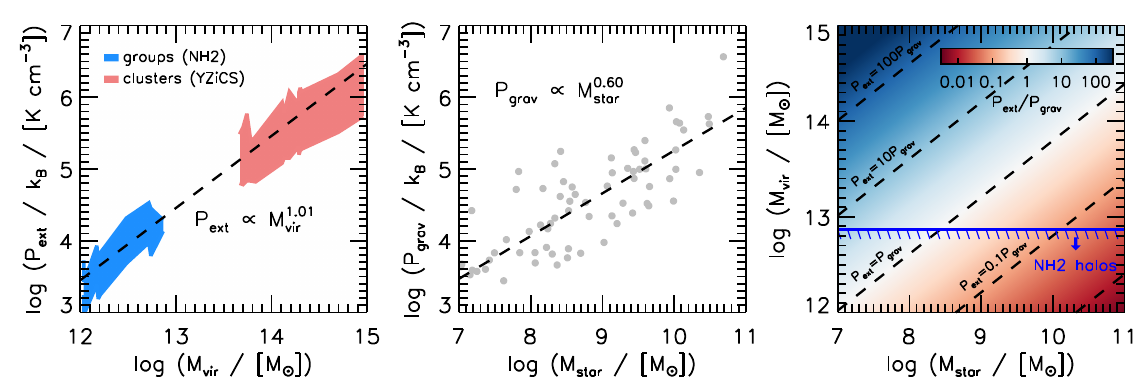}
\caption{
Left panel: Ram pressure as a function of halo virial mass.
The ram pressure is estimated using representative values of gas density and galaxy velocity within $0.2-0.5\,R_{\rm vir}$ of each halo.
The blue shaded region shows the range of ram pressures measured from the \NHtwo\ group halos, while the red shaded area represents those from cluster halos in \cite{CY17}.
The dashed line represents a power-law fit to the halos, yielding $P_{\rm ext} \propto\,M_{\rm vir}^{1.01}$.
Middle panel: Gravitational pressure of normal satellite galaxies as a function of their stellar mass (gray circles).
The power-law fit to the normal galaxies (black dashed line) is $P_{\rm grav} \propto M_{*}^{0.6}$, with a lower power index than the external pressure of halos.
Right panel: Ratio of external to gravitational pressure using the fits on the stellar mass-halo mass plane.
The background color represents this pressure ratio; the bluer regions indicate that external pressure exceeds gravitational pressure.
The upper bound of the virial mass of the \NHtwo\ halos is indicated by the blue solid line.
In these halos, satellite galaxies with $M_{*}\lesssim 10^{8.4}\,M_{\odot}$ may experience external pressure that exceeds their gravitational restoring force.
} 
\label{fig:fig massdep}
\end{figure*}

This galaxy has recently passed its apocenter in Panel-(A) and begins moving toward the pericenter between Panels (B) and (D), and thus, a gas tail is observed pointing toward the group center.
In Panel (B), when the galaxy approaches its pericenter, a nearby galaxy exhibits a prominent outflow, highlighted by the sharp layer with the negative velocity divergence of gas components.
This outflow propagates outward, and by Panel (C), its leading edge reaches the target galaxy.
The insets in Panel (C) capture this interaction in detail: the temperature and density map show a clear discontinuity at the wake front.
By Panel (D), the outflow continues to impinge on the galaxy, and the galaxy thus begins to penetrate into the outflow bubble, leading to significant gas stripping.
The hot gas in the bubble generally has a very high speed, $\sim 500 \,{\rm km\,s^{-1}}$, which far exceeds the speed corresponding to the virial velocity within the host halo.
This is evidenced by the disappearance of the central gas clump in the density inset, marking the onset of gas deficiency.
Therefore, the strong ram pressure acting on this galaxy originates from the outflow of a nearby galaxy rather than from the ambient IGM.
This type of physically driven ram pressure can be strong regardless of the orbital velocity of galaxies.
Thus, we conclude that a stochastic physical event, such as an outflow from a nearby galaxy, can be an important driver of strong ram pressure---responsible for gas stripping in satellite galaxies.
This also demonstrates that group halos display complex environmental effects on their satellites.

\subsection{Mass dependence of gas stripping}
\label{sec:Discussion-Mass}

One of the notable features of group environments is the presence of a critical stellar mass threshold, below which SF quenching becomes significantly efficient.
In particular, galaxies with $M_* < 10^{8}$ are frequently quenched, while more massive systems typically remain star forming in Local Group analogs \citep[e.g.,][]{Wheeler14, Fillingham15, Akins21, Rhee24}.
This trend is also evident in the gas mass-stellar mass scaling relation shown in Fig.~\ref{fig:fig scaling}, where gas-deficient satellites are predominantly low-mass galaxies.
To investigate the physical origin of this critical mass, we assume that gas stripping is the primary quenching mechanism and examine how it depends on both stellar and host halo mass.

The left panel of Fig.~\ref{fig:fig massdep} presents the characteristic ram pressure acting on galaxies as a function of their host virial mass.
For each halo, the ram pressure is estimated as $\rho\, v^{2}$, where $\rho$ is the hot-phase gas density in the halo (defined following \cite{Torrey12}) and $v$ is the escape velocity as a proxy for free-fall velocities of satellite galaxies, estimated using Equation (1) from \cite{Rhee17}, assuming a concentration index of 6.
These quantities are evaluated within the radial range of $0.2-0.5\,R_{\rm vir}$ to represent typical conditions experienced by infalling satellites.
The blue-shaded region corresponds to the ram pressure values from the \NHtwo\ halos, while the red-shaded region shows the corresponding measurements from cluster-scale halos in the \yzics\ simulation \citep[][]{CY17}.
A clear positive correlation between ram pressure and halo mass is evident.
A single power-law fit yields $P_{\rm ext} \propto M_{\rm vir}^{1.01}$, implying a nearly linear scaling of ram pressure with halo mass.
The slope is slightly higher than the theoretically predicted value of 2/3 in Equation 17 in \cite{Boselli22}, perhaps due to the additional dependence of the gas density on halo mass.

The middle panel of Fig.~\ref{fig:fig massdep} shows the gravitational restoring pressure of normal satellite galaxies as a function of stellar mass.
Gas-deficient satellites are excluded from this analysis, as the measurement of $P_{\rm grav}$ requires the definition of $R_{\rm 90, gas}$, which cannot be determined for galaxies without gas.
A power-law fit is $P_{\rm grav} \propto M_{*}^{0.60}$, indicating that gravitational pressure scales with stellar mass more weakly than external pressure does with halo mass.
In contrast, when gravitational pressure is fit against the dynamical mass within $R_{\rm 90, gas}$, the relation becomes nearly linear: $P_{\rm grav}\propto M_{\rm dyn}^{1.07}$.
The relatively shallow slope with stellar mass implies that stellar mass is not a direct proxy for the dynamical mass of galaxies responsible for the restoring force.
For instance, in galaxies with $M_{*}<10^{9}\,M_{\odot}$, the gas fraction can exceed unity (see Fig.~\ref{fig:fig scaling}), meaning that gas constitutes a significant portion of their dynamical mass.
As stellar mass increases, the gas fraction tends to decline, and the stellar component progressively dominates the total dynamical mass.
Therefore, stellar mass alone may serve as a biased estimator of the restoring force in galaxies.

In the right panel of Fig.~\ref{fig:fig massdep}, the ratio of external to gravitational pressure is shown on the stellar mass-halo mass plane, based on the power-law fits to both quantities.
The background color represents the pressure ratio ($P_{\rm ext}/P_{\rm grav}$), with bluer regions indicating regimes where external pressure dominates.
The guidelines with $P_{\rm ext} = 100, 10, 1, 0.1,$ and $0.01\, P_{\rm grav}$ are drawn with the black dahsed lines. 
The blue solid line marks the upper bound of the virial mass of the \NHtwo\ halos.
Galaxies with $M_{*} \gtrsim 10^{8.4} \, M_{\odot}$ in the \NHtwo\ halos lie in regions where $P_{\rm grav} > P_{\rm ext}$, consistent with the prevalence of gas-rich satellites in this mass regime in this paper.
Conversely, the lower-mass satellites are exposed to stronger external pressure relative to their gravitational restoring pressure, increasing their vulnerability to gas stripping.
These results suggest that the emergence of the critical mass for efficient gas stripping in low-mass group halos observed in various studies can be attributed to the intersection point of the scaling relations between external pressure and gravitational pressure with halo mass and stellar mass, respectively, which occurs around $M_{*}\sim10^{8}\,M_{\odot}$ in low-mass group halos.
In addition, in a typical cluster mass with $M_{\rm vir}\sim 5\times10^{14}\,M_{\odot}$, galaxies with $M_{*}<10^{11.5}\,M_{\odot}$ lie in the external pressure-dominant zone, likely explaining the higher frequency of jellyfish galaxies in cluster \citep[e.g.,][]{Roberts21}.
The corresponding mass ratio, $M_{\rm vir}/M_{*} \sim 3$, also coincides with the range in which environmental effects become a dominant driver of star formation quenching in cluster environments \citep[][]{Jeon22}.

This analysis provides only a rough estimate of the critical stellar mass for gas stripping and SF quenching.
Thus, we admit that a more rigorous statistical framework could yield a different threshold, owing to several complicating factors.
As discussed in Sect.~\ref{sec:Discussion-RP}, ram pressure likely operates stochastically, with transient structures such as nearby outflows or clumps in the IGM generating localized enhancements in pressure \citep[see also][]{Fillingham16, Simons20, Samuel23}.
These events can result in a broad diversity of gas stripping outcomes among otherwise similar galaxies.
Second, the gravitational restoring force is not static.
It can dynamically respond to the external ram pressure---ISM compression by ram pressure increases the gravitational restoring pressure.
This effect is visible in the median gravitational pressure of satellite galaxies at the time when the external pressure reaches the peak in Fig.~\ref{fig:fig pressure} (blue shaded region), which is generally higher than the corresponding values shown in the middle panel of Fig.~\ref{fig:fig massdep}.
In addition, galaxies can have different orbital parameters, affecting the exposure to ram pressure.
Galaxies on circular orbits tend to experience weaker ram pressure than those on radial trajectories \citep[e.g.,][]{Jung18}, a factor not incorporated in Fig.~\ref{fig:fig massdep}.
Lastly, while Fig.~\ref{fig:fig massdep} implies that a significant fraction of galaxies with $10^{7-8}\,M_{\odot}$ can retain their gas in Local Group-like halos ($M_{\rm vir}\sim(1-3)\times 10^{12}\,M_{\odot}$), observations show that most galaxies in this mass range are gas poor in the actual Local Group.
This discrepancy may be explained by the dynamical state of the Local Group, specifically the ongoing merger between the MW and M31 halos.
Cluster mergers are known to amplify ram pressure \citep[e.g.,][]{Roediger14}, and similarly, paired systems like the Local Group may experience enhanced stripping \citep[e.g.,][]{Samuel23}.
This may partially reconcile the mismatch between the predictions from Fig.~\ref{fig:fig massdep} and empirical data from the Local Group \citep[][]{McConnachie12,Putman21}.

The complexity increases further when considering the role of a critical mass in SF quenching.
While some low-mass galaxies may retain their gas in the presence of ram pressure, their SF can still be quenched by alternative mechanisms \citep[][]{Emerick16, Fillingham16}.
For example, \cite{Rhee24} demonstrated that interactions between the hot IGM and the galactic ISM can reduce SF efficiency by altering the phases of ISM.
Conversely, even in cases of gas stripping, the central cold-phase ISM may remain intact, allowing continued SF activity \citep[e.g.,][]{Hausammann19, Cortese21}.
Thus, the existence of a critical stellar mass---whether for gas stripping or SF quenching---may stem from the different scaling relations between external pressure on halo mass and gravitational pressure on stellar mass.
However, a range of physical processes also contribute to the diversity and scatter observed in galaxy populations.
A more statistically robust analysis is needed to disentangle these effects, which will be addressed in future work.

\subsection{Effects from tidal stripping}
\label{sec:Discussion-Tidal}

\begin{figure*}
\centering
\includegraphics[width=0.95\textwidth]{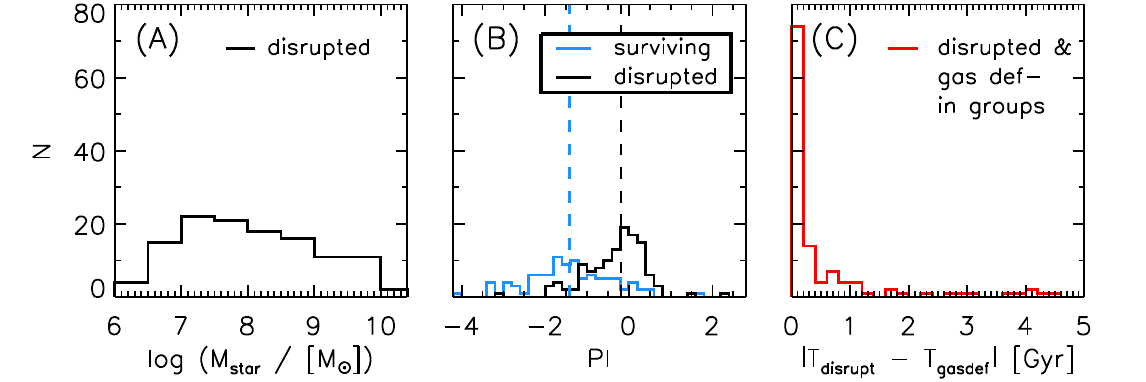}
\caption{
Panel (A): Stellar mass distribution of disrupted satellite galaxies.
The maximum stellar mass of the disrupted galaxies is used in the histogram.
Panel (B): Distributions of perturbation index for surviving (blue) and disrupted (black) satellites.
The dashed vertical lines indicate the median of each distribution in the corresponding color.
The disrupted galaxies show remarkably higher perturbation indices than the surviving ones, suggesting that they are disrupted by interactions with nearby objects within their host groups.
Panel (C): Time difference between the epoch of gas loss and the epoch of disruption for each disrupted galaxy.
The marginal time difference suggests that the gas loss and gravitational perturbations occurred nearly simultaneously.
}
\label{fig:fig tidal}
\end{figure*}

In Sect.~\ref{sec:Results-Def}, we focused on the role of hydrodynamic pressure in driving gas removal, motivated by the fact that all gas-deficient galaxies show strong external pressure at the time of stripping (Fig.~\ref{fig:fig prcompare}).
However, the low velocity dispersion in group halos increases the chance of galaxy-galaxy encounters compared to clusters, potentially amplifying the impact of gravitational tidal forces \citep[][]{Yi13}.
This is consistent with observations of disturbed \HI\ morphology in satellite galaxies within low virial mass systems \citep[e.g.,][]{Putman98,Pearson16, Luber22}.

To assess the impact of gravitational tides, we begin by revisiting the definition of galaxy gas mass adopted in this study.
Our analysis focuses on the gas content within the effective radius (see Sect.~\ref{sec:Data-Gas}), which may neglect the influence of tidal forces acting beyond this radius.
Moreover, the steep radial gradient of gravitational pressure within the effective radius, evident in the radial profiles of Fig.~\ref{fig:fig himap}, suggests that the ISM is strongly concentrated toward the galaxy centers.
Accordingly, gas removal in this work refers to the complete loss of the ISM, including its centrally concentrated reservoir.
Having this idea on hand, if tidal forces were the dominant mechanism driving gas removal, they would also be expected to perturb other bound mass components within the effective radius \citep[e.g.,][]{Mayer06, Marasco16, Iorio19}.
Therefore, galaxies that have experienced strong tidal stripping are unlikely to retain their stellar or DM cores by the final epoch.

We identified 119 additional galaxies that were satellites of the host groups at $z<1$, but are no longer detected in the final snapshot.
Figure~\ref{fig:fig tidal} presents the properties of these ``disrupted galaxies.''
Panel-(A) displays their maximum stellar mass distribution before disruption.
Although these galaxies have disappeared before $z=0.158$, they are not confined to the low-mass systems but instead span a broad range of stellar masses.
A detailed, galaxy-by-galaxy investigation reveals that many of the disrupted satellites were destroyed through interactions with neighboring galaxies.

This raises the question of why a substantial number of galaxies become disrupted while others remain intact.
To quantify the cumulative perturbative influence experienced by satellite galaxies following their infall into host groups, we adopt the perturbation index (PI) \citep[][]{BV90, Choi18}, defined as 
\begin{equation}
{\rm PI} = \log \int_{t_{1}}^{t_{2}}dt / [{\rm Gyr}] \,\sum_{i} \left(\frac{M_{*,i}}{M_{*,0}}\right)\left(\frac{R_{\rm eff}}{d_{i}}\right)^3,
\end{equation}
where $M_{\rm *,0}$ and $M_{*}$ are the stellar mass of the target and $i$-th nearby galaxy, respectively, and $R_{\rm eff}$ and $d_{\rm i}$ are the half-mass radius of the target galaxy and the distance to the $i$-th galaxy.
The PI is computed over a time interval from infall to the final snapshot.
Panel-(B) of Fig.~\ref{fig:fig tidal} shows the PI distributions for the disrupted galaxies (black) and the surviving satellites (blue), with the dashed lines indicating their respective medians.
The disrupted population shows systematically higher PI values, indicating that these galaxies have undergone significantly stronger cumulative interactions with neighboring galaxies.
It is likely that such perturbative encounters are the primary mechanism responsible for their disruption.
Additionally, tidal forces from global group potential, which are not accounted for in the PI calculation, may further contribute to the structural transformation and eventual disruption of these satellites.

Panel-(C) shows the time interval between the gas removal and the disruption of each galaxy.
Here, we include only the disrupted galaxies that were gas rich at the infall moment.
In most cases, gas loss occurs within $\lesssim 1\,{\rm Gyr}$ before disruption, suggesting that gas stripping coincides with or is a part of the perturbation-driven destruction process.
These results imply that gravitational interactions, particularly close encounters with neighboring galaxies, not only affect the gas content but can also lead to the complete structural disruption of satellites.
When such interactions are sufficiently strong to influence the centrally concentrated ISM, they typically affect the entire galaxy, ultimately making it unobservable at late times.
Consequently, restricting the analysis to surviving systems may underestimate the true impact of gravitational encounters on gas stripping of galaxies in group environments.

\begin{figure}
\centering
\includegraphics[width=0.48\textwidth]{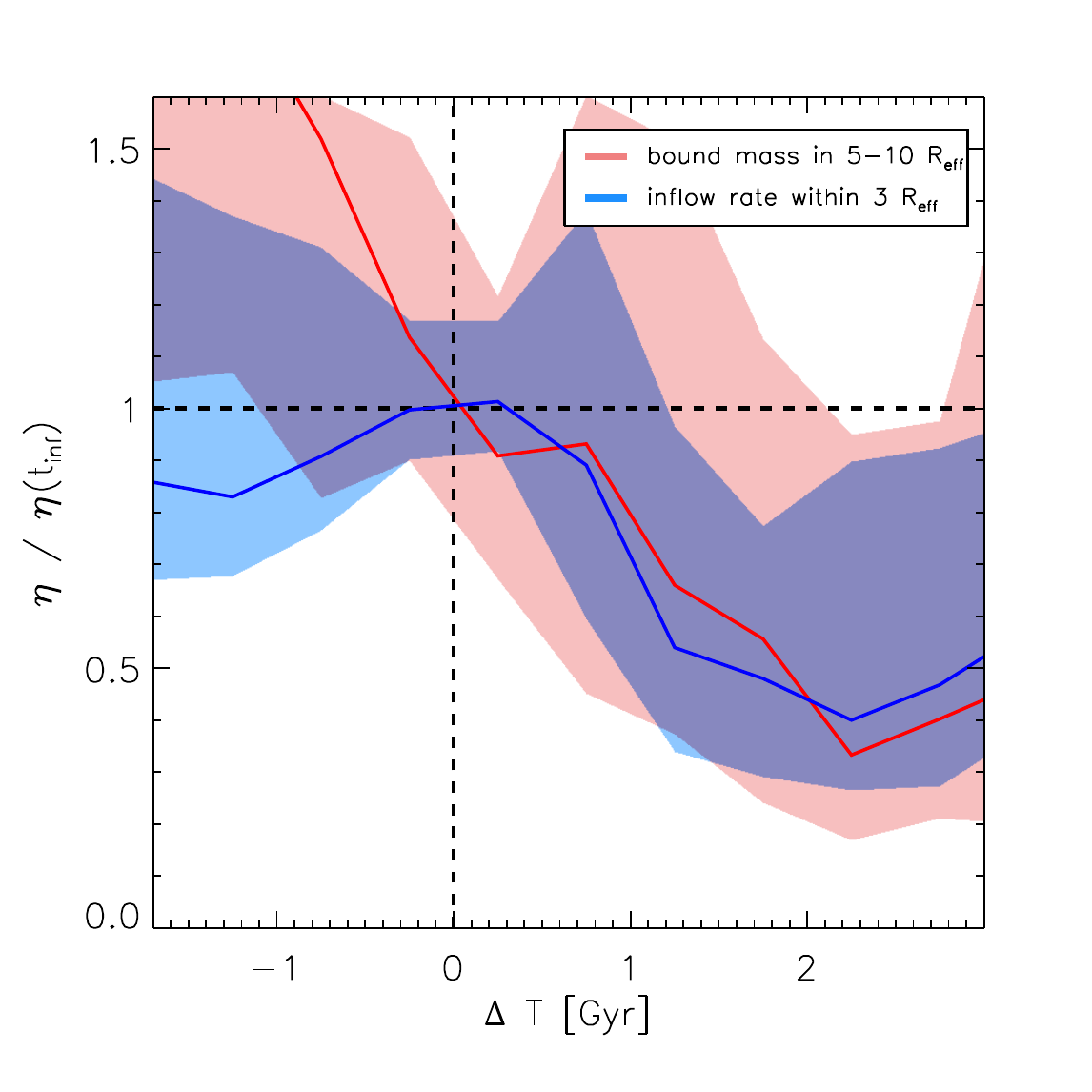}
\caption{
Time evolution of bound gas mass in the outer region ($5-10\,R_{\rm eff}$) and the inflow gas rate into the inner region $3\,R_{\rm eff}$ for normal satellites.
The evolution of each galaxy is time-aligned such that infall occurs at $\Delta T = 0$.
All quantities are normalized to their respective values at the time of infall.
The red and blue solid lines show the median evolution of bound mass and gas inflow rate of the galaxies, respectively, while the shaded regions represent the interquartile ranges.
Both quantities decrease after infall, reaching $30-40\,\%$ of their initial values at infall, indicating substantial gas stripping in the outer gaseous halos of normal satellites.
}
\label{fig:fig starvation}
\end{figure}

\subsection{Starvation}

Normal satellite galaxies in Fig.~\ref{fig:fig gmassevol} show minimal changes in their gas mass over time.
Given their substantial gas reservoirs, they are capable of sustaining SF over extended timescales.
A plausible quenching mechanism for them is starvation \citep[][]{Fillingham15, Fillingham16, Akins21}, wherein the inflow of fresh cold gas is suppressed after infall, and SF gradually declines as the remaining internal gas is consumed.

We investigate whether this starvation scenario applies to the \NHtwo\ satellite galaxies.
Figure~\ref{fig:fig starvation} shows the time evolution of the bound gas mass of normal satellite galaxies in the $5-10\,R_{\rm eff}$ range, which traces hot halo gas of individual satellites, and the gas inflow rate into $3\,R_{\rm eff}$.
The gas inflow rate is measured by the total mass of gas tracer particles that transitioned from beyond $3\,R_{\rm eff}$ to within $3\,R_{\rm eff}$ between two consecutive snapshots.
To evaluate the relative differences in these quantities after infall, each quantity is normalized to its value at the time of infall.
As predicted by the starvation scenario \citep[][]{Larson80}, the halo gas of normal satellite galaxies, of which the cold ISM remains largely unaffected, undergoes substantial stripping, losing approximately $60-70\,\%$ of its mass, by the time of infall.
This loosely bound halo gas is therefore susceptible to removal within the group environment.
In parallel, the gas inflow into $3\,R_{\rm eff}$ also diminishes significantly.
These results indicate that normal satellite galaxies experience a significant reduction in external gas accretion following their infall.

However, the normal satellite galaxies are unlikely to be quenched by this mechanism.
For instance, the median depletion timescale for normal satellite galaxies, defined as the ratio of ISM gas mass to the SFR, substantially exceeds the Hubble time.
This timescale may be even longer when considering that SF in satellite galaxies is often episodic \citep[e.g.,][]{Rhee24}.
This is partly consistent with long SF quenching timescales via starvation in several studies \citep[][]{Fillingham16}.
Therefore, while galaxies lose a significant amount of hot gas after infall, starvation process appears to play a minimal role in quenching ongoing SF.

\section{Summary and Conclusions}

In this paper, we investigate the physical mechanisms responsible for gas removal in satellite galaxies ($M_{*}>10^{7}\,M_{\odot}$) in low-mass group halos ($10^{12-13}\,M_{\rm vir}$), using the \NHtwo\ simulation.
We select nine group halos and a total of 100 satellite galaxies, including nine BGGs from the final snapshot with $z=0.158$.
Among these satellites, 71 retain substantial gas reservoirs at the final snapshot, resistant to complete gas stripping (Fig.~\ref{fig:fig scaling}).
These gas-rich galaxies show nearly constant gas mass over time (the upper panel of Fig.~\ref{fig:fig gmassevol}), suggesting that their gas reservoirs are largely unaffected by environmental processes throughout their evolution.
However, when the gravitational restoring force in the outer regions falls below the surrounding external pressure, these galaxies can develop asymmetric gas morphologies, particularly near the group center.
This results in jellyfish-like features (Fig.~\ref{fig:fig himap}), which may be detectable in group environments by future high-sensitivity \HI\ surveys with facilities such as the SKAO.

In contrast, 20 galaxies have low gas content ($f_{\rm gas}< 0.01$), and these are predominantly low-mass galaxies ($M_* \lesssim 10^{8}\,M_{\odot}$).
This indicates that a substantial fraction of low-mass satellite galaxies are gas deficient, in agreement with previous studies reporting a high prevalence of gas-poor galaxies in this mass regime \citep[e.g.,][]{Fillingham16}.
Among the 20 gas-deficient galaxies at the final snapshot, 12 were already gas poor prior to infall, and one galaxy lost its gas due to strong internal feedback, suggesting no clear link to host group environmental processes in these cases.
The remaining seven galaxies underwent complete gas stripping within their host groups.
These galaxies experience rapid gas loss that occurs near their pericenter passages (the bottom panel of Fig.~\ref{fig:fig gmassevol}).
These galaxies experience external pressure comparable to, or exceeding, their gravitational restoring force at the time of gas loss (Fig.~\ref{fig:fig prcompare}).
The source of this strong external pressure includes both conventional ram pressure from hot IGM-ISM interactions and fast encounters with cold ambient gas (Fig.~\ref{fig:fig igmphase}).
These findings reveal that gas removal in group environments can arise through a broader range of mechanisms than previously recognized.

Moreover, we find that external pressure does not necessarily lead to complete gas stripping.
As shown in Fig.~\ref{fig:fig prcompare}, Galaxy ID=236 undergoes three distinct episodes of strong external pressure, triggered either by an interaction with another galaxy or by moderately strong ram pressure.
At the same time, the gravitational restoring force at the galaxy center increases as the ISM is compressed in response to the external pressure.
Thus, these events cause a temporary redistribution of the ISM but do not fully strip the gas.
In contrast, complete gas removal is associated with exceptionally strong external pressure---significantly exceeding both that experienced by other satellites and the restoring force of the galaxies (Fig.~\ref{fig:fig pressure}).
This suggests that such extreme conditions likely arise from stochastic physical processes (e.g., galaxy-galaxy interaction).
In Fig.~\ref{fig:fig rporigin}, one galaxy provides a clear example in which a neighboring galactic outflow drives strong external pressure, leading to a high-speed collision between the ISM and hot gas accelerated by the outflow.

The emergence of a critical stellar mass for efficient gas stripping (or SF quenching), around $\sim 10^{8}\,M_{\odot}$, appears to stem from the different scaling behaviors of external pressure and gravitational restoring pressure with halo and stellar mass, respectively (see Fig.~\ref{fig:fig massdep}).
External pressure increases more rapidly with halo mass than the gravitational restoring force does with stellar mass.
This likely reflects the fact that the total dynamical mass responsible for the gravitational restoring force does not scale linearly with stellar mass; for instance, the gas fraction tends to decline with increasing stellar mass.
As a result, a crossover between external and gravitational restoring pressure emerges around $10^{8}\,M_{\odot}$ in low-mass group systems ($M_{\rm vir}\sim10^{12-13}\,M_{\odot}$), suggesting the existence of a critical mass scale.
Nonetheless, this threshold is not sharply defined, as it is subject to scatter introduced by variations in halo dynamical states, stochasticity in external pressure, and orbital diversity among galaxies.

Although this study has focused on hydrodynamic pressure, tidal forces induced by galaxy interactions appear to be an important mechanism for gas removal.
The low velocity dispersion characteristic of group environments increases the likelihood of close galaxy-galaxy encounters.
Indeed, a substantial fraction of satellite galaxies are found to be disrupted within their host groups, and these disrupted galaxies show systematically higher perturbation indices compared to the surviving satellites (Panel (B) of Fig.~\ref{fig:fig tidal}).
This suggests that frequent and strong gravitational interactions with neighboring galaxies can play a major role in satellite disruption.
Moreover, these disrupted galaxies tend to lose their gas at the time of their structural disruption (Panel (C) of Fig.~\ref{fig:fig tidal}).
Therefore, tidal forces sufficiently strong to completely remove the ISM also disrupt the overall stellar and DM structure of the galaxy.
As a consequence, the role of tidal interactions in gas removal may be underrepresented in studies that consider only surviving galaxy populations.

Finally, we assessed the validity of the starvation quenching scenario for gas-rich satellite galaxies (Fig.~\ref{fig:fig starvation}).
Although these galaxies retain a nearly constant ISM gas mass over time, they show significant loss of their outer halo gas following infall into the host group.
Simultaneously, the infall rate of gas onto the central regions of the galaxies declines.
Therefore, the first requirement of the starvation scenario---namely, the shutdown of cold gas inflow---is satisfied.
However, the depletion timescale for these galaxies, defined as the ratio of ISM gas mass to SFR, is found to exceed the Hubble time, likely due to a substantial cold gas reservoir that remains intact.
Furthermore, as galaxies in group halos can have low SFR during their evolution, their actual depletion timescales may be even longer than the instantaneous estimates.
We therefore conclude that the starvation scenario is an inefficient mechanism for quenching SF in these gas-rich satellite galaxies with $M_* \sim 10^{8-10}\,M_{\odot}$.

Effects from environments are an important factor in regulating the evolutionary paths of galaxies.
Among them, gas stripping is widely recognized as a primary channel driving SF quenching in satellite galaxies.
In clusters, both observations and numerical simulations have established a consensus that gas stripping is a dominant process regulating the gas content and subsequent SF activity of galaxies.
However, its role in lower-mass systems remains less explored, in part due to observational challenges associated with detecting faint galaxies in these environments.
The relatively shallow gravitational potentials of low-mass group halos raise the question of whether gas stripping is still an effective mechanism in such regimes.
To address this, we analyze satellite galaxies in the \NHtwo\ simulation and find that gas stripping remains an important environmental process even in group scale halos ($M_{\rm vir}\sim 10^{12-13}\,M_{\odot}$), although it manifests differently than in clusters.
The weaker ram pressure in groups gives rise to a broader range of galactic responses, reflecting the interplay between environmental forces and internal galaxy structure.
We admit that the statistical significance of our results is limited by the sample size of satellite galaxies.
Future studies with large samples of group systems will be essential for establishing a more comprehensive and statistically robust understanding of satellite galaxy evolution in low-mass systems.

\begin{acknowledgements}
We thank Dirk Scholte and Yao-Yuan Mao for sharing the data for Fig.~\ref{fig:fig scaling}.

J.R. was supported by the KASI-Yonsei Postdoctoral Fellowship and was supported by the Korea Astronomy and Space Science Institute under the R\&D program (Project No. 2025-1-831-02), supervised by the Korea AeroSpace Administration.
This work was partially supported by the Institut de Physique des deux infinis of Sorbonne Universit\'{e} and by the ANR grant ANR-19-CE31-0017 of the French Agence Nationale de la Recherche.
S.K.Y acknowledges support from the Korean National Research Foundation (RS-2025-00514475; RS-2022-NR070872).

This work was granted access to the HPC resources of
CINES under the allocations c2016047637, A0020407637,
and A0070402192 by Genci; KSC-2017-G2-0003, KSC2020-CRE-0055, and KSC-2020-CRE-0280 by KISTI; and as
a ``Grand Challenge'' project granted by GENCI on the AMD
Rome extension of the Joliot Curie supercomputer at TGCC.
The large data transfer was supported by KREONET, which is managed and operated by KISTI.

W.L. acknowledges support from the National Research Foundation of Korea (NRF) grant funded by the Korea government (MSIT) (RS-2024-00340949).

E.C. and S.J. acknowledge support from the Korean National Research
Foundation (RS-2023-00241934).

B.L. acknowledges support by the National Research Foundation of Korea, grant Nos. RS-2022-NR069020.

J.L. is supported by the NRF of Korea grant funded by the Korea government (MSIT, RS-2021-NR061998). J.L. also acknowledges the support of the NRF of Korea grant funded by the Korea government (RS-2022-NR068800).
\end{acknowledgements}

\bibliographystyle{aa}
\bibliography{ref}

\end{document}